\documentclass[tighten, twocolumn]{aastex631}
\usepackage{lineno}
\usepackage{CJK}

\shorttitle{The Tilted \& Doubly Broken Stellar Halo}
\shortauthors{Han et al.}

\graphicspath{{./}{figures/}}

\begin{document}
\begin{CJK*}{UTF8}{gbsn}

\title{The Stellar Halo of the Galaxy is Tilted \& Doubly Broken}

\correspondingauthor{Jiwon Jesse Han}
\email{jesse.han@cfa.harvard.edu}

\author[0000-0002-6800-5778]{Jiwon Jesse Han}
\affiliation{Center for Astrophysics $|$ Harvard \& Smithsonian, 60 Garden Street, Cambridge, MA 02138, USA}

\author[0000-0002-1590-8551]{Charlie Conroy}
\affiliation{Center for Astrophysics $|$ Harvard \& Smithsonian, 60 Garden Street, Cambridge, MA 02138, USA}

\author[0000-0002-9280-7594]{Benjamin D. Johnson}
\affiliation{Center for Astrophysics $|$ Harvard \& Smithsonian, 60 Garden Street, Cambridge, MA 02138, USA}

\author[0000-0003-2573-9832]{Joshua S. Speagle (\begin{CJK*}{UTF8}{gbsn}沈佳士\ignorespacesafterend\end{CJK*})}
\affiliation{Department of Statistical Sciences, University of Toronto, 9th Floor, Ontario Power Building, 700 University Ave, Toronto, ON M5G 1Z5, Canada}
\affiliation{David A. Dunlap Department of Astronomy \& Astrophysics, University of Toronto, 50 St George Street, Toronto ON M5S 3H4, Canada}
\affiliation{Dunlap Institute for Astronomy \& Astrophysics, University of Toronto, 50 St George Street, Toronto, ON M5S 3H4, Canada}
\affiliation{Data Sciences Institute, University of Toronto, 17th Floor, Ontario Power Building, 700 University Ave, Toronto, ON M5G 1Z5, Canada}

\author[0000-0002-7846-9787]{Ana Bonaca}
\affiliation{The Observatories of the Carnegie Institution for Science, 813 Santa Barbara St., Pasadena, CA 91101, USA}

\author[0000-0002-0572-8012]{Vedant Chandra}
\affiliation{Center for Astrophysics $|$ Harvard \& Smithsonian, 60 Garden Street, Cambridge, MA 02138, USA}

\author[0000-0003-3997-5705]{Rohan P. Naidu}
\affiliation{Center for Astrophysics $|$ Harvard \& Smithsonian, 60 Garden Street, Cambridge, MA 02138, USA}

\author[0000-0001-5082-9536]{Yuan-Sen Ting (丁源森)}
\affiliation{Research School of Astronomy \& Astrophysics, Australian National University, Cotter Road, Weston Creek, ACT 2611, Canberra, Australia}
\affiliation{Research School of Computer Science, Australian National University, Acton ACT 2601, Australia}

\author[0000-0002-0721-6715]{Turner Woody}
\affiliation{Center for Astrophysics $|$ Harvard \& Smithsonian, 60 Garden Street, Cambridge, MA 02138, USA}

\author[0000-0002-5177-727X]{Dennis Zaritsky}
\affiliation{Steward Observatory, University of Arizona, 933 North Cherry Avenue, Tucson, AZ 85721-0065, USA}

\begin{abstract}

Modern Galactic surveys have revealed an ancient merger that dominates the stellar halo of our Galaxy (\textit{Gaia}-Sausage-Enceladus, GSE). Using chemical abundances and kinematics from the H3 Survey, we identify 5559 halo stars from this merger in the radial range $r_{\text{Gal}}=6-60\text{ kpc}$. We forward model the full selection function of H3 to infer the density profile of this accreted component of the stellar halo. We consider a general ellipsoid with principal axes allowed to rotate with respect to the Galactocentric axes, coupled with a multiply-broken power law. The best-fit model is a triaxial ellipsoid (axes ratios 10:8:7) tilted $25^\circ$ above the Galactic plane towards the Sun and a doubly-broken power law with breaking radii at 12 kpc and 28 kpc. This result resolves the long-standing dichotomy in literature values of the halo breaking radius, being at either $\sim15\text{ kpc}$ or $\sim30\text{ kpc}$ assuming a singly-broken power law. N-body simulations suggest that the breaking radii are connected to apocenter pile-ups of stellar orbits, and so the observed double-break provides new insight into the initial conditions and evolution of the GSE merger. Furthermore, the tilt and triaxiality of the stellar halo could imply that a fraction of the underlying dark matter halo is also tilted and triaxial. This has important implications for dynamical mass modeling of the Galaxy as well as direct dark matter detection experiments.

\end{abstract}

\section{Introduction} \label{sec:intro}

Accounting for $\sim1\%$ of the total stellar mass, the stellar halo plays an outsized role in our understanding of the Galaxy. Since the foundational works of \citet{ELS62} and \citet{SZ78}, it has been realized that the halo is teeming with archaeological ``fossils'' in the form of stellar overdensities, streams, and clusters in integrals of motion \citep[see reviews by e.g., ][]{freeham20,belokurov13,gerhard-BH16, johnston16,helmi20}. These fossils string together a story of how hierarchical structure formation assembled our Galaxy, through which we can learn about the physics of galaxy formation and the nature of dark matter.

The field of Galactic Astronomy has been revolutionized with the \textit{Gaia} mission \citep[][]{gaia16}. The 6D phase space of nearby stars revealed that the local halo (distance $d\sim2\text{ kpc}$) is composed of stars on highly radial orbits and a kinematically heated disk component \citep[e.g.,][]{bonaca17, belokurov18, Helmi18, haywood18, Gaia18a, belokurov20}. Subsequent studies expanded this result beyond the local halo, revealing that the bulk of halo stars are indeed on eccentric orbits and follow a well-defined sequence in the [Fe/H]---$[\alpha/\text{Fe}]$ plane \citep[e.g.,][]{mackereth19,naidu20}, clearly distinguishable from the thick disk of the Galaxy. This $[\alpha/\text{Fe}]$-poor eccentric population of the halo, \textit{Gaia}-Sausage-Enceladus \citep{belokurov18,Helmi18}, is interpreted to be the debris of a major merger of a dwarf galaxy with the Milky Way progenitor at redshift $z=1-2$ \citep[e.g.,][]{grand18,mackereth18, gallart19, bonaca20, belokurov20}. The GSE merger likely gave rise to the bulk of the accreted stellar halo and dynamically heated the existing disk of the Galaxy into the in-situ halo \citep[e.g.,][]{bonaca17, belokurov20}. The timeline of this scenario is well supported by several lines of evidence, notably from the ages of stars measured by various techniques \citep[e.g.,][]{gallart19, bonaca20, belokurov20, chaplin20, borre21, Grunblatt21, montalban21}. While the accreted origin of the Galactic halo has been studied before \citep[e.g.,][]{bell08}, the new insight brought by GSE is that the inner $\sim30$ kpc halo was largely built from \textit{one} accretion event.

Parallel to efforts that have charted out substructures of the stellar halo, studying the overall density profile of the halo has also been an active area of research \citep[e.g.,][]{watkins09,sesar11,Deason11,sesar13,faccioli14, deason14,pilz-diez15,xue15,H+18}. The radial density profile reflects the cumulative accretion history of the Galaxy, and is most often approximated by a singly-broken power law. The Sagittarius stream \citep[][]{ibata01, newberg02,Majewski03, belokurov06} is often excluded or treated separately from this measurement, as it contributes significant mass in the outer halo in the form of a kinematically cold and spatially coherent stream. The measured slope of the power law density profile can then be used to compare to stellar halos from simulations \citep{BJ05,pillepich14,pillepich18}. Various tracers have been used to measure the radial density profile, including near-main sequence turnoff, blue horizontal branch (BHB), RR-Lyrae (RRL), and red giant branch stars. Each tracer is sensitive to different radial ranges and stellar populations (e.g., RRL and BHB stars are preferentially old and metal poor). A census of previous measurements of the radial density profile is presented in \cite{gerhard-BH16}. Many of these studies share the same underlying model, which is an oblate spheroid that is aligned with the Galactic disk (i.e., flattened radius $r_q\equiv \sqrt{X^2+(Y/p)^2+(Z/q)^2}$, where $p=1$ is fixed) and a single-break power law. There is however considerable spread in the best-fit profiles reported in the literature. For example, the flattening parameter $q$ ranges from $0.59$ \citep{Deason11} to near-spherical \citep{H+18}, the outer power law slope from $2.7$ \citep{sesar13} to $5.8$ \citep{sesar10}, and the breaking radius from $16$ kpc \citep{sesar13} to $34.6$ kpc \citep{sesar10}. Such a spread in the measured density profile could suggest that the adopted model of an oblate spheroid and a single-break power law does not encapsulate the true underlying distribution, thus yielding different shape parameters along different sightlines. \citet{Lowing15} show that this is the case for halos in the Aquarius simulation, which are rich in both unrelaxed and fully phase-mixed substructure. It is now known that the Milky Way halo is also filled with substructure, and one of them dominates the stellar mass: GSE.

A connection between the density profile and GSE was made by \citet{deason18}. They find metal-rich, highly-eccentric halo stars (likely belonging to GSE) that have a common apocenter at 20 kpc, roughly coinciding with the literature values of the radial density breaking radius. From this they argue that the break in the radial density profile is due to the apocentric pile-up of a massive accretion event. Soon after, \citet[][]{simion19} showed that two major stellar overdensities, the Hercules-Aquila Cloud \citep[HAC; ][]{HAC} and the Virgo Overdensity \citep[VOD;][]{VOD, juric08}, are linked to GSE based on their similar integrals of motion. Using \textit{Gaia} DR1 cross-matched with 2MASS \citep{skrutskie06}, \citet[][]{ib18} show that the inner ($r_{\text{Gal}}<28\text{ kpc}$) stellar halo occupies a triaxial ellipsoid (axes ratio $10:7.7:5.0$) that is rotated with respect to the Sun-Galactic Center axis by $70^\circ$. Using this model, \citet{IB19} show that the halo is bookended by HAC and VOC, bolstering the argument made by \citet{simion19} that these overdensities are components of GSE that have not been fully phase mixed in the Galaxy. They also speculate on a possible tilt with respect to the Galactic plane. \citet[][N21]{naidu21} ran a grid of $\sim500$ N-Body galaxy mergers to identify a plausible configuration of GSE, based on the radial density and angular momentum distribution of the present day halo. From the fiducial simulation, they find that GSE occupies a strongly triaxial ellipsoid (axes ratios 10:7.9:4.5) with its major axis misaligned with the Sun-Galactic Center axis by $20^\circ$ (consistent with \citealt{IB19}) and tilted away from the Galactic plane by $35^\circ$. In addition, they find that the radial profile has two breaking points, corresponding to two apocenter passages of GSE before it fully merged with the Galaxy. These studies demonstrate two points. First, GSE has a profound influence on the shape and density profile of the halo. Second, an oblate spheroid model may be insufficient to describe GSE.

\begin{figure}
    \centering
    \includegraphics[width=0.45\textwidth]{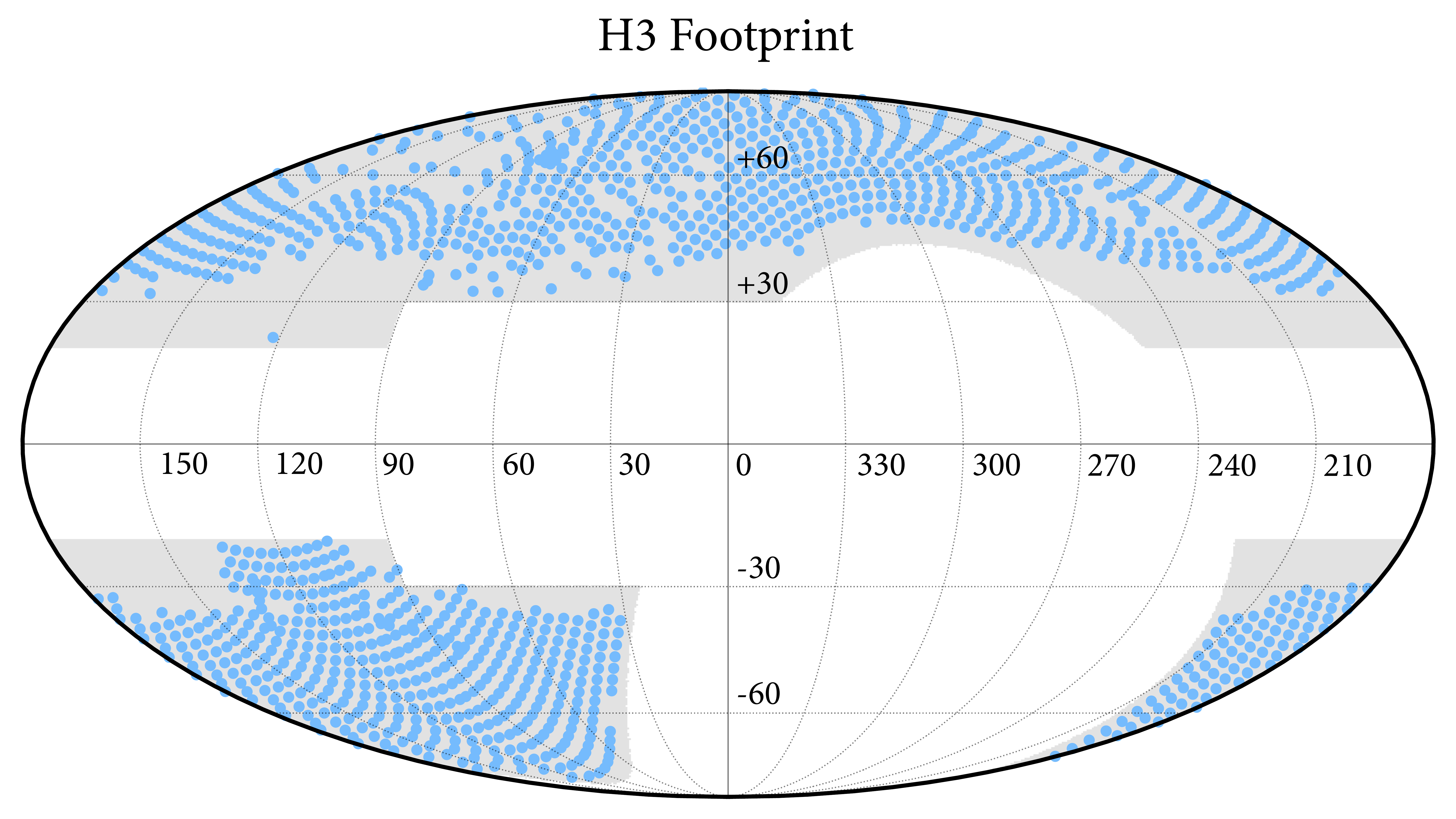}
    \caption{H3 Survey sky coverage as of May 2022 in Galactic coordinates. Fields observed are shown in blue symbols (not drawn to scale). The white region delineates $\delta=-20^\circ$ and $|b|=20^\circ$ within $180^\circ$ of the Galactic anticenter ($90^\circ<l<270^\circ$), and $|b|=30^\circ$ within $180^\circ$ of the Galactic center ($-90^\circ<l<90^\circ$). Following figures showing on-sky projections follow the same coordinate conventions.}
    \label{fig:H3 footprint}
\end{figure}

\begin{figure*}
    \centering
    \includegraphics[width=0.96\textwidth]{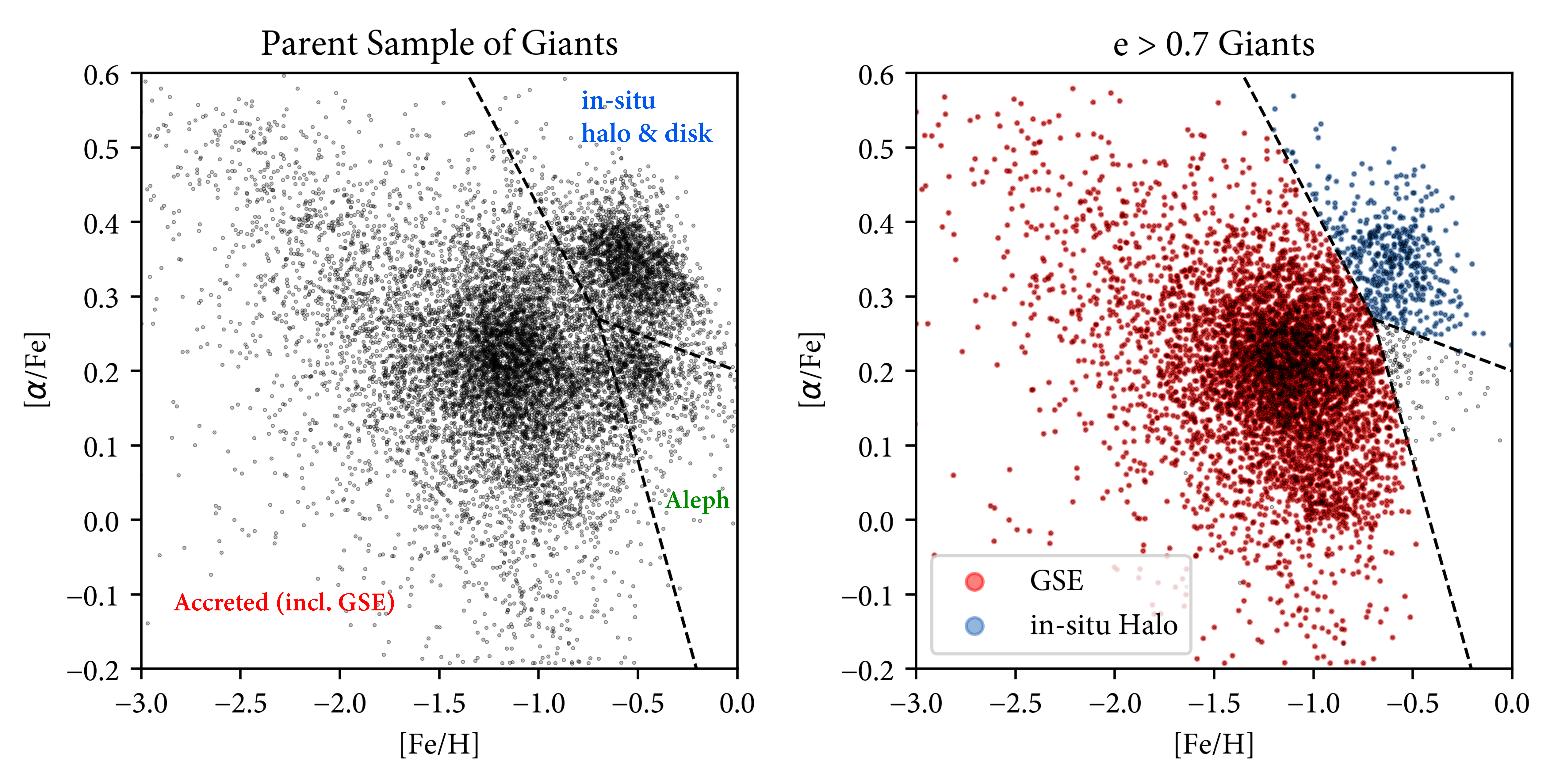}
    \caption{H3 giants in [Fe/H] - $[\alpha/\text{Fe}]$ space. Kinematically cold structures such as the Sagittarius stream and dwarf galaxies have been removed. In the left panel, the three most prominent clumps are the in-situ (``high-$\alpha$'') disk/halo, GSE, and Aleph. On the right panel, we plot stars that are on eccentric ($e>0.7$) orbits from the parent sample. We can easily identify GSE (overplotted in red) and the in-situ halo (overplotted in blue). GSE accounts for 90\% of the $e>0.7$ sample.}
    \label{fig:chemspace}
\end{figure*}

\begin{figure}
    \centering
    \includegraphics[width=.48\textwidth]{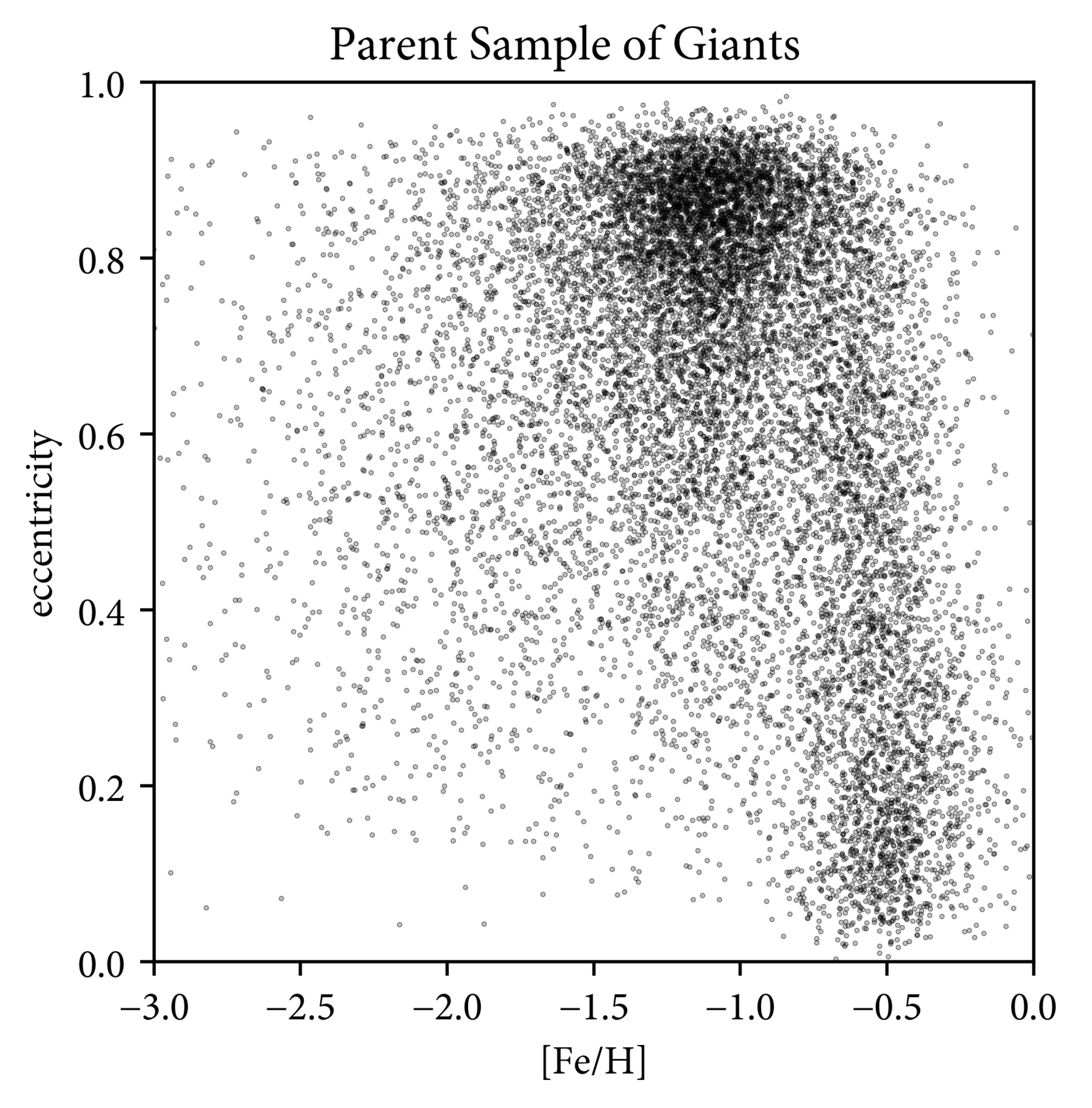}
    \caption{Eccentricity vs. [Fe/H] for the parent sample of giants. GSE occupies the high-$e$ clump at $\text{[Fe/H]}\lesssim -1$ and the thick disk and in-situ halo occupy the low eccentricity strip at $\text{[Fe/H]}\sim-0.5$. GSE clearly dominates the sample at high $e$, and its distribution extends down to $e\approx0.5$.}
    \label{fig:espace}
\end{figure}

Despite these significant advancements in our understanding of the halo, the shape and density profile of GSE have yet to be well constrained. A major challenge is that the adequate data has been lacking. To identify GSE stars amid thick disk and in-situ halo stars, one needs to know the 6D phase space coordinates and chemical abundances. In particular, the abundance of $\alpha$-elements has been shown to be effective in distinguishing accreted halo populations from the in-situ population \citep[e.g.,][]{mackereth19,naidu20, das20}. Both the kinematic and chemical data has been limited, as \textit{Gaia} does not deliver reliable radial velocities outside of the local halo ($d\lesssim5\text{ kpc}$). Even for the farther-reaching RRL stars to which we have reliable distances, \textit{Gaia} does not provide accurate radial velocities or elemental abundances; thus, it is difficult to distinguish between disk and halo stars and, for halo stars, between in-situ and accreted components.  Furthermore, \cite{balbinot21} raise a concern that the faint limit of the \textit{Gaia} RRL may be affected by \textit{Gaia}'s scanning law, which could affect the observed global distribution of RRL on the scales of $\sim30\text{ kpc}$. 

For these reasons, extensive spectroscopic data is crucial to supplement \textit{Gaia} to enable chemo-dynamical analysis of the stellar halo and GSE. For example, \citet{mackereth20} cross-match \textit{Gaia} DR2 with APOGEE DR14 \citep{majewski17, abolfathi18} to obtain mono-abundance samples in $\text{[Fe/H]}-\text{[Mg/Fe]}$ space to infer the stellar mass of GSE. However, with a few hundred stars identified to be part of GSE (out of 835 total number of giants), the data was not sufficient to constrain the full density profile. \citet{wu22} cross match \textit{Gaia}-EDR3 \citep{EDR3} and K-giants from LAMOST DR5 \citep{zhao06,liu15} to study the density profile of GSE, but they rely on [Fe/H] and 3D velocities to identify GSE stars, which makes it difficult to distinguish between GSE and in-situ halo stars. Furthermore, they fit an oblate spheroid with a single power law in line with previous studies of the stellar halo, which limits the investigation of any spatial anisotropy in the data.

The H3 Survey \citep{conroy19a} provides high-resolution spectroscopy to complement \textit{Gaia}. By combining the two datasets, we can obtain full 6D phase space information, [Fe/H], and $[\alpha/\text{Fe]}$ for a quarter million stars, and parse through the various substructures of the halo in chemo-dynamical space and identify which stars likely belong to GSE.

We thus set out to measure the shape and density profile GSE - the dominant component of the accreted stellar halo - with the H3 Survey. Specifically, we aim to test two salient features of the N21 model that have not been explored in previous models of the stellar halo: an overall tilt of the halo with respect to the disk, and a doubly-broken power law along its principal axes.

\section{The H3 Survey} \label{sec:sample}

The H3 Survey \citep[][]{conroy19a} is collecting spectra of 300,000 stars in high Galactic latitude fields using the MMT telescope with the medium-resolution Hectoschelle spectrograph at R=32,000 \citep[][]{Szentgyorgyi11} over the wavelength range $5150$\AA$-5300$\AA. Figure \ref{fig:H3 footprint} shows the survey footprint. As of May 2022, H3 has observed 1242 tiles and 238,372 stars, covering $\delta>-20^\circ$ and $|b|>20^\circ$ within $180^\circ$ of the Galactic anticenter ($90^\circ<l<270^\circ$), and $|b|>30^\circ$ within $180^\circ$ of the Galactic center ($-90^\circ<l<90^\circ$). One of the most notable features of the H3 survey is its simple selection function. H3 targets stars from Pan-STARRS \citep{chambers16}, primarily based on \textit{Gaia} parallax and $r$-band magnitude. \textit{Gaia} DR2 was not available early on in the survey, so a subset of the targets were selected on $g-r<1.0$ instead. These stars only comprise 10\% of the sample, and the results in this study do not change if the early data are excluded. A detailed summary of the H3 selection function is presented in the Appendix. Based on this selection function, we describe how we model selection effects in the Methods section.

From H3 spectra, stellar parameters are derived using the \texttt{MINESweeper} program \citep{Cargile20} including radial velocities, spectrophotometric distances, iron abundances [Fe/H], and $\alpha$ element abundances [$\alpha$/Fe]. \texttt{MINESweeper} also infers initial chemical abundances, but for giants these two values are nearly identical and for this study we use surface chemical abundances. Combined with \textit{Gaia} proper motions, these quantities can then be converted to orbital quantities such as angular momentum, eccentricity, $e$, and energy using \texttt{gala} \citep{gala:joss, gala} assuming the default \texttt{MWPotential} \citep[disk model from][]{bovy15}.

The stellar halo as observed through the H3 Survey is rich in substructure. A comprehensive cataloguing of these structures is carried out in N20. Here, we focus on identifying GSE, the single most dominant contributor to the accreted stellar halo.

To obtain a pure halo sample, we first select giants ($\log g < 3.5$) that do not have any data quality flags and $\text{SNR}>5$. The giant selection filters out the nearby dwarf stars, and the SNR cut ensures that chemical abundances are sufficiently precise to separate GSE from other populations. We then remove any cold structures from this data set. The major contaminant is the Sagittarius stream; H3 identifies stars that belong to the Sagittarius stream based on angular momenta \citep{johnson20} and we can remove them from the sample. Similarly, we remove dwarf galaxies, globular clusters, and kinematically cold stellar streams from our sample (Sextans, Ursa Minor, Palomar 13, Draco, M31, M33, M92, and GD-1) based on their 6D phase space coordinates and chemistry. We define the remaining giants to be the parent sample.

Figure \ref{fig:chemspace} shows the parent sample in $[\alpha/\text{Fe}]-[\text{Fe/H}]$ space. In the left panel we show the whole parent sample, revealing three conspicuous clumps. These clumps are GSE (low [Fe/H] and low $[\alpha/\text{Fe}]$), the in-situ disk and halo (high [Fe/H] and high $[\alpha/\text{Fe}]$), and Aleph (high [Fe/H] and low $[\alpha/\text{Fe}]$, likely of in-situ origin). Owing to the prominence of these clumps, we can simply draw lines in the $[\alpha/\text{Fe}]-[\text{Fe/H}]$ plane to cleanly select on each structure (dashed lines). In the right panel, we show giants that are on eccentric ($e>0.7$) orbits. The two prominent clumps are GSE (overplotted in red) and the in-situ halo (overplotted in blue). GSE accounts for more than 90\% of the $e>0.7$ sample, demonstrating the dominance of GSE in the stellar halo. A small fraction of the high-$\alpha$ spur at low [Fe/H] can be attributed to an in-situ population, but this structure accounts for less than $<5\%$ of the inner halo \citep{belokurov22, conroy22}, and does not affect the density profile of the accreted halo in any significant way.

Figure \ref{fig:espace} shows the parent sample in $\text{[Fe/H]}-\text{eccentricity}$ space. The prominent clump at high eccentricities ($e>0.7$) is GSE, and the $\text{[Fe/H]}\sim-0.5$ strip that extends from low to high eccentricities includes the in-situ halo, thick disk, and Aleph. We see that the eccentricity distribution of GSE tapers off around $e\sim0.7$, and extends down to $e\sim0.5$.

Combining information from Figure \ref{fig:chemspace} and Figure \ref{fig:espace}, we define GSE as the following:

\begin{itemize}
    \item[]$e>0.7$ and
    \item[]$[\alpha/\text{Fe}] < 0.27-0.5(\text{[Fe/H]}+0.7)$ and
    \item[]$[\alpha/\text{Fe}] < 0.27-0.1(\text{[Fe/H]}+0.7)$.
\end{itemize}

As a result, we obtain 5559 stars with a high probability of belonging to GSE. This selection criterion is intended to maximize the purity of the GSE sample rather than its completeness, because our goal is to measure the shape of the accreted halo against the background of the in-situ halo and the thick disk. We discuss variations to the fiducial sample in section \ref{sec:results}. We note that the choice of $e>0.7$ is common in the literature \citep[e.g.,][]{mackereth20,naidu20,wu22}.

\begin{figure*}
    \centering
    \includegraphics[width=\textwidth]{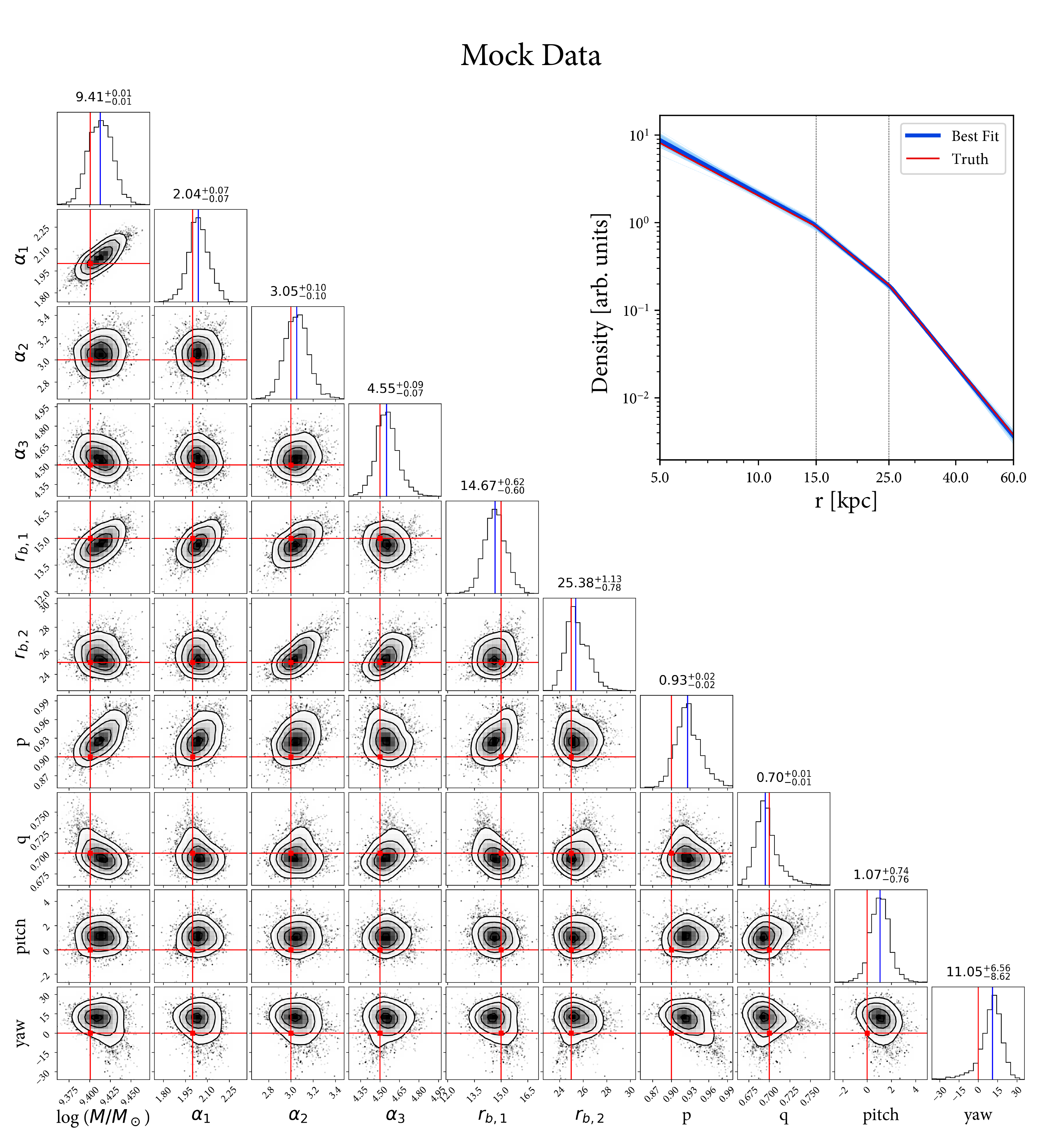}
    \caption{Posterior distribution from fitting a mock data set constructed to resemble the H3 Survey. Blue lines indicate the sample mode, and red lines are the true value. The true parameters are all within the statistical uncertainty of the fit. We adopt a uniform prior for all fits. Breaking radii are in units of kpc.}
    \label{fig:mock corner}
\end{figure*}

\begin{figure}
    \centering
    \includegraphics[width=.5\textwidth]{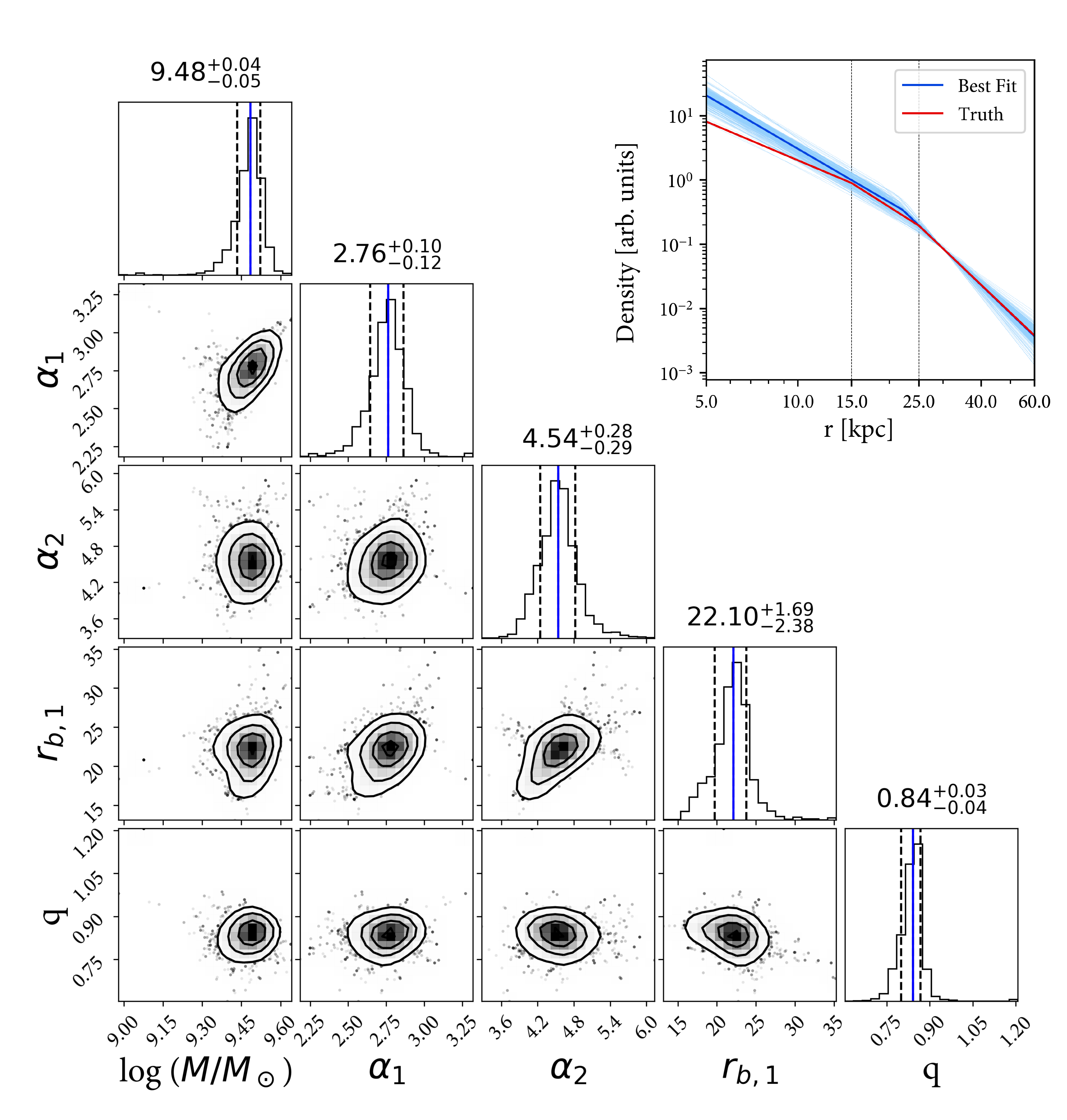}
    \caption{Posterior distribution from fitting the mock data in Figure \ref{fig:mock corner} with a restricted model.  Here we allow for one breaking radius (units in kpc) and one flattening parameter, and no rotations. This is akin to models of the stellar halo in the literature. Blue lines indicate the sample mode, and dashed lines indicate $16^{th}$, $84^{th}$ percentiles. Despite the the posterior being strongly constrained to be an oblate spheroid that is aligned with the Galactic disk, the underlying model is triaxial and tilted, and model residuals should be clear when projected onto data space.}
    \label{fig:5d mock}
\end{figure}

\section{Methods} \label{sec:methods}

In this Section we outline how we infer the stellar density profile of GSE based on H3 data.

\subsection{Likelihood Function}

We model each star as drawn from an inhomogeneous Poisson point process. This likelihood-based method to constrain density profiles was pioneered by studies such as \citet[][]{bovy12, rix13, xue15}. We define the fundamental variables that determine the Poisson state of a star to be the 3D location $\vec{r}$, initial mass $m$, and metallicity [Fe/H]. We express $\vec{r}$ in either Galactocentric coordinates ($X_{gal}$, $Y_{gal}$, $Z_{gal}$) or Heliocentric coordinates ($l$, $b$, $d$). While stellar age is another fundamental variable, most GSE stars are within a stellar age range of $8-12\text{ Gyr}$ \citep[e.g.,][]{gallart19, bonaca20}. Hence, stellar ages do not significantly affect the observable parameters, and we assume a uniform stellar age of 10 Gyr for this study. The intrinsic distribution of the state variables, the \textit{rate function} $\lambda$, is assumed to be separable as follows:
\begin{equation}
    \lambda = \rho(\vec{r}) \cdot \text{IMF}(m) \cdot \text{MDF}(\text{[Fe/H]}).
\end{equation}

\noindent
The assumption of separability is equivalent to assuming that the MDF and IMF do not depend on Galactic location. This assumption is supported by the lack of a metallicity gradient in the halo over the range $6<r_{\text{Gal}}<100\text{ kpc}$, measured by \citet[][]{conroy19b}. We adopt a Kroupa initial mass function \citep[IMF;][]{kroupa01} and an empirical metallicity distribution function (MDF) derived from a fitted accretion model \citep[see][for details]{conroy19b}. Folding in the H3 selection function, the observed rate can be written as:

\begin{equation}
    \lambda_{obs} = \lambda(\vec{r},m,\text{Fe/H}) \cdot S(\text{tile}),
\end{equation}
where $S$ is a tile-dependent selection function on observed stellar photometry and \textit{Gaia} parallax $\pi$. To transform the fundamental variables $m$ and [Fe/H] into photometry, we use \texttt{MIST} v1.2 isochrones \citep[][]{MIST0,MIST1} at 10 Gyr and place them at a heliocentric distance $d$ derived from $\vec{r}$. The H3 target selection involves a cut on $r$-band magnitude, and the limiting magnitude of our giant selection for a given tile further depends on factors such as sky brightness and weather. A histogram of limiting magnitude for SNR=5 is presented in the Appendix, and we take this into account in our selection function. While $S$ should intuitively be a binary function (a star is either observable or not observable), it is actually a continuous function from 0 to 1. First, \textit{Gaia} parallaxes have a statistical uncertainty that causes stars at the edge of detectability to have non-negligible scatter. We account for this effect by marginalizing the rate function over the \textit{Gaia} parallax uncertainty, which is a function of the ecliptic angle, $G$ band magnitude, and $V-I$ color. Second, we can only place a maximum of 200 fibers within any given field; sometimes even fewer than 200 to avoid fiber collisions among the robotic positioners. If there are more targets than fibers available, we must include the sampling fraction (defined as $N_{\text{received fiber}}/N_{\text{target}}$) in the observed rate function. This sampling fraction is calculated based on the parent Pan-STARRS catalog, and we include it in the rate function as a constant multiplicative factor for each tile. Finally, H3 implements a ranked targeting scheme, where rank 1 stars (BHB, RRL, photometrically-selected luminous K-giants) are given priority over rank 2 stars (selected on parallax and $r$ magnitude). Thus, the sampling fraction is also a function of the rank of the target. We make note that rank 1 stars only comprise 6\% of the sample. Accounting for the ranks, the observed rate function can be written as:

\begin{equation}
    \lambda_{obs} = \lambda(\vec{r},m,\text{[Fe/H]}) \cdot
    \left [ \mathcal{B}\cdot f_{r_1} + \mathcal{B}\cdot f_{r_2} \right],
\end{equation}
where $\mathcal{B}$ is a Boolean function on the state variables that determine if the star is rank 1, rank 2, or non-observable; $f_{r_1}$ and $f_{r_2}$ are sampling fractions of the respective target ranks. The definition of rank has evolved over the duration of the survey, and we report these changes in the Appendix. We note that there also exist rank 3 (faint filler targets) and rank 4 (nearby filler targets), but these targets comprise less than 1\% of the giants with SNR$>5$ and we exclude them from our analysis.

Once we have specified the observed rate function, we can evaluate the likelihood function for a Poisson process:

\begin{equation}
    \log \mathcal{L} = -N_{\lambda} + \sum_{i} \lambda_{obs}(\mathcal{D}_i),
\end{equation}
where $\mathcal{D}_i \equiv ( \vec{r}_i, m_i, \text{[Fe/H]}_i )$, and 
\begin{equation}
    N_\lambda = \int \lambda_{obs} \text{ d}\vec{r} \text{ d}m \text{ d}\text{[Fe/H]}.
\end{equation}
Intuitively, $N_\lambda$ represents the expected number of events (stars) observed through the selection function given a set of model parameters. While this integral would be computationally formidable to perform over the full five-dimensional space (three spatial dimensions, $m$, and metallicity), it is tractable if we approximate all of the stars in one tile to lie along one line-of-sight, and thus collapse three spatial dimensions to one. Since the size of each tile ($0.5^\circ$ radius) is much smaller than the total footprint, we confidently make this approximation.

We now discuss the functional form of $\rho(\vec{r})$. Assuming that the halo can be approximated by an ellipsoid (which includes spherical and flattened spherical shapes), there are two parts to consider in parameterizing $\rho(\vec{r})$. The first is the radial profile, representing the density along the principal axes of the ellipsoid. Previous studies have fit a singly-broken power law or a Sersic profile to stellar halo tracers. Here, we consider a multiply-broken power law that encompasses a no-break to doubly-broken power law.

The second component of $\rho(\vec{r})$ is the angular profile, representing the overall orientation of the ellipsoid. This profile has historically not been explored as in depth as the radial profile. However, there is no a priori reason to assume that a merger remnant would occupy an azimuthally symmetric distribution in the Galaxy; indeed, there is increasing evidence to consider a halo that is allowed to be rotated with respect to the Galactic plane \citep[][]{IB19,n21,han22}. In this study, we consider the simplest form of the angular profile, which is to assume the ellipsoid as a solid body that has been rotated. More flexible models could consider the rotation angles changing as a function of radius, and we leave this to future work. Combining the radial and angular profiles, we write the 3D density profile as follows:

\begin{figure*}[h]
    \centering
    \includegraphics[width=\textwidth]{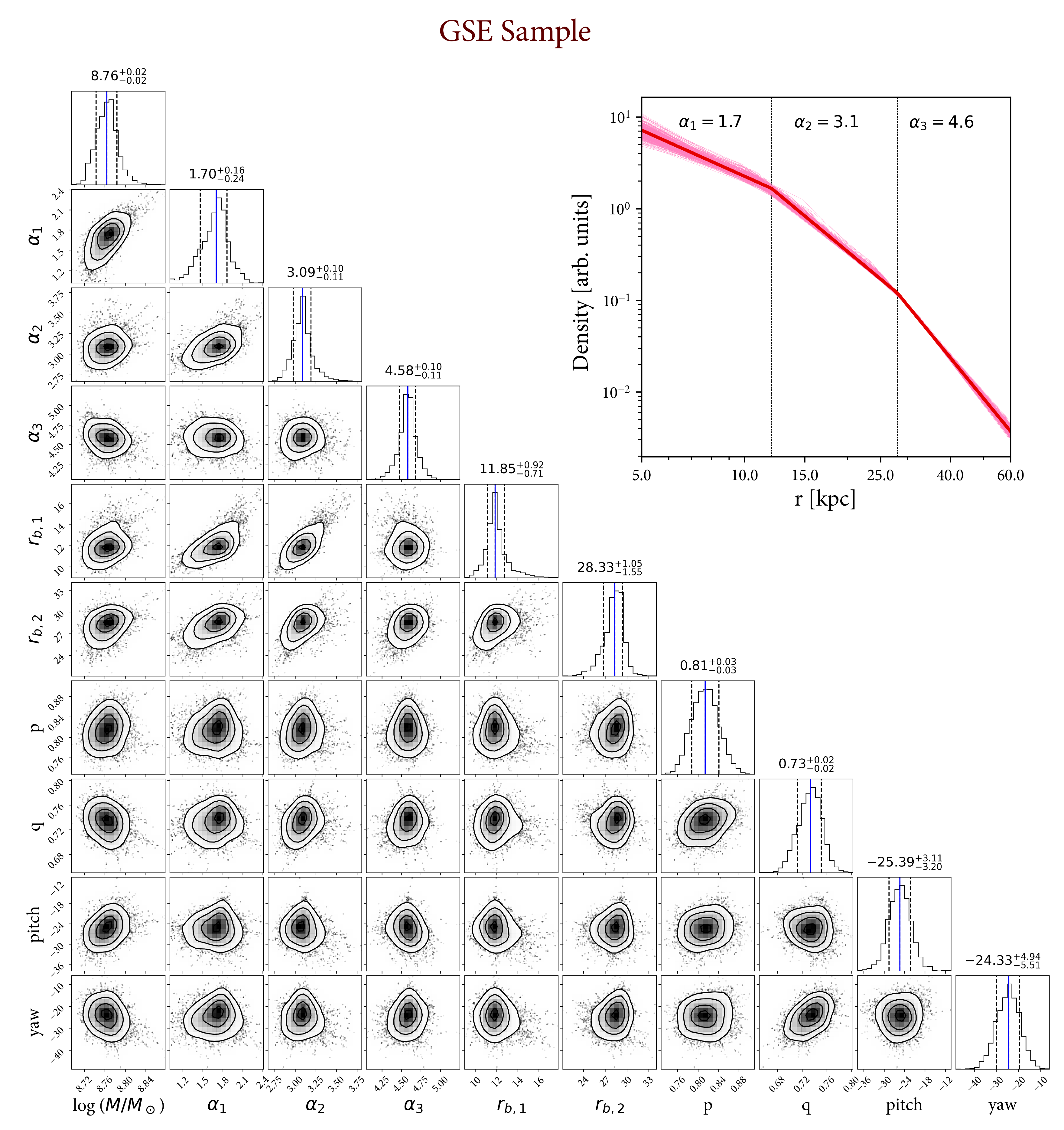}
    \caption{Posterior distribution of the fiducial model applied to the GSE sample. Blue lines indicate the sample mode, and dashed lines indicate $16^{th}$, $84^{th}$ percentiles. The covariances among various parameters are mostly small, suggesting that our model parameters are non-degenerate with one another. To the upper right, we plot the radial density profile based on the best-fit parameters, highlighting the double break feature of the power law.}
    \label{fig:10d corner}
\end{figure*}
\begin{figure*}[]
    \centering
    \includegraphics[width=.85\textwidth]{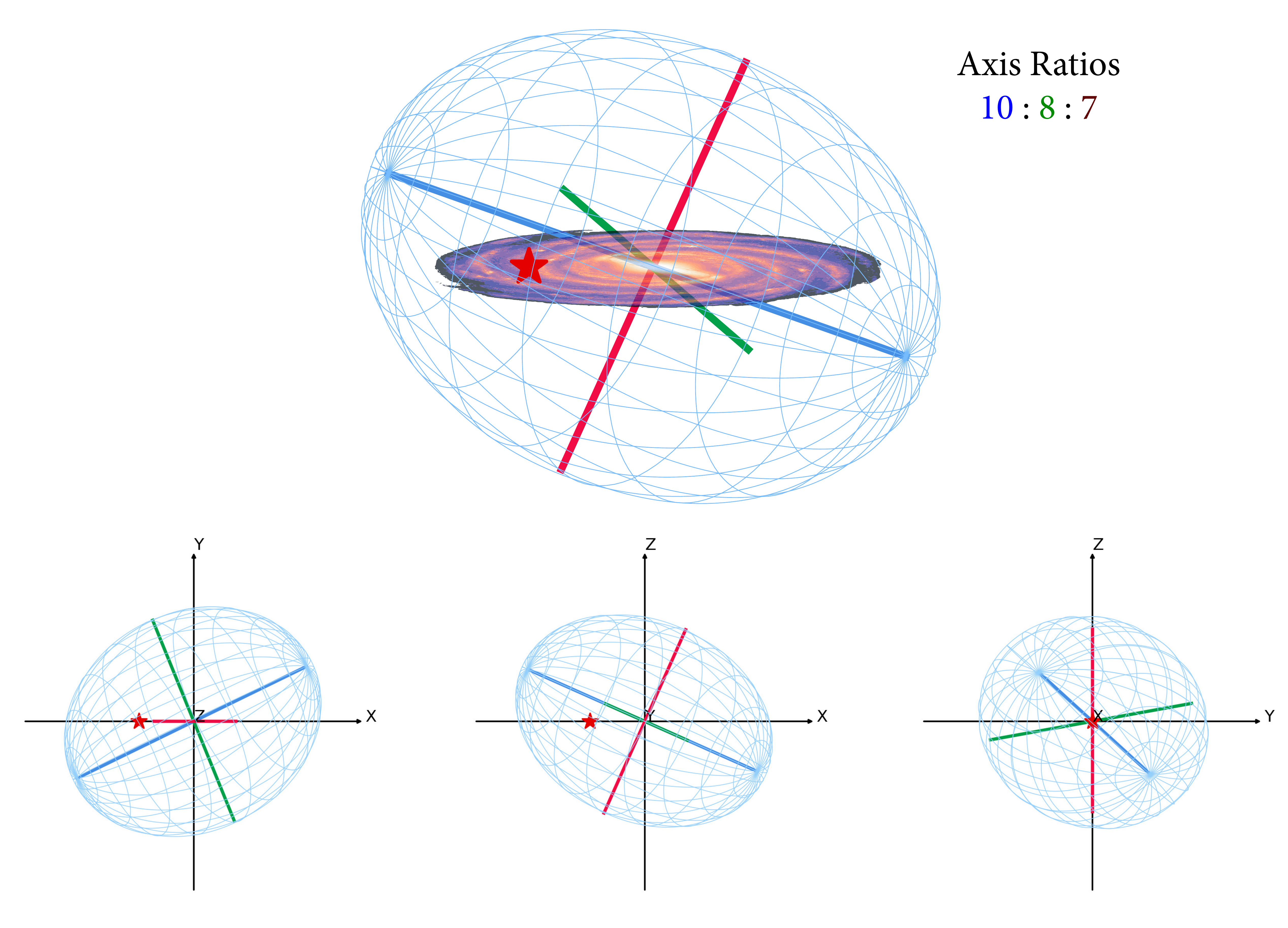}
    \caption{Isodensity surface contours of the fiducial ellipsoid model evaluated at flattened radius $r_q=20\text{ kpc}$. The major, intermediate, and minor axes of the ellipsoid are respectively drawn in blue, red, and green. The solar location is marked with a red star, displaced $8$ kpc away from the Galactic center. An artist impression of the Galactic disk \citep{MWdisk} is overplotted to orient the reader. The principal axes ratios are 10:8:7, which is closer to prolate than oblate. The major axis (blue) is tilted $25^\circ$ with respect to the Galactic plane, and also $24^\circ$ with respect to the Sun-Galactic Center axis.}
    \label{fig:fiducial model}
\end{figure*}

\begin{equation}
    \rho(\vec{r}) = f(M,\alpha_1,\alpha_2,\alpha_3,r_{b,1},r_{b,2},p,q,\text{pitch}, \text{yaw}).
\end{equation}

\noindent
Here, $f$ is a multiply-broken power law that is described by ten parameters: total stellar mass $M$, first power law slope $\alpha_1$, second power law slope $\alpha_2$, third power law slope  $\alpha_3$, first breaking radius $r_{b1}$, second breaking radius $r_{b2}$, intermediate axis flattening $p$, minor axis flattening $q$, pitch angle, and yaw angle. The power law is a function of the flattened radius $r_q\equiv \sqrt{X^2+(Y/p)^2+(Z/q)^2}$, where $p$ and $q$ are values between $0$ and $1$. The power law slopes and breaking radii are allowed to overlap such that $f$ can describe a single power law to a triple power law (i.e. doubly-broken). The \textit{pitch} angle represents the angle of the major axis of the ellipsoid with respect to the Galactic plane (colloquially referred to as \textit{tilt}), and the \textit{yaw} angle represents the azimuthal angle of the major axis with respect to the Galactic Z-axis. As we discuss in Section \ref{sec:results}, the best-fit ellipsoid model is near-prolate, and thus has approximate internal azimuthal symmetry. Hence, we do not fit for the rotation of the ellipsoid about the long axis, i.e. \textit{roll} angle. 

Finally, to write the density function in terms of observed coordinates $(l, b, d)$, we must account for the change in the volume element as a function of observed coordinates. This is the absolute value of the Jacobian determinant, and can be derived as $||J||(l, b, d|\vec{r})=d^2\cos(b)$.

The task at hand of computing the likelihood function can naturally be parallelized over tiles. We use OpenMPI \citep{openmpi} to distribute the likelihood function over $1000$ CPUs, enabling on-the-fly calculations of the likelihood instead of having to rely on precomputed grids. We assume a uniform prior for all parameters, and use \texttt{emcee} \citep[][]{FM13} to sample from the posterior.

\subsection{Mock Tests}
The forward model framework described above is also a generative model with which we can create mock data sets. Given true model parameters, we can sample stars from the spatial density, IMF, and MDF. We can then pass these stars through the full H3 selection function to create a mock data set. Such mock data sets are useful for several purposes. First, we can fit the mock data set through our full fitting pipeline to validate our inference framework.  An example of such a fit is displayed in Figure \ref{fig:mock corner}. The diagonal panels show the posterior of each parameter as sampled with \texttt{emcee}, while other panels show the covariance among different parameters. The blue lines indicate the sample mode (also written as text on top of the diagonal panels), and the red lines indicate the true parameters that were inputs to the mock data. All of the true parameters are well within the statistical uncertainties of the inferred parameters, demonstrating the robustness of our inference framework. 

Second, we use mock catalog to understand how model assumptions can affect our interpretation of the data. For example, we create a tilted, triaxial, doubly-broken mock data set (from the same parameters as Figure \ref{fig:mock corner}), and then fit it with a non-tilted, spheroidal (one flattening parameter $q$), singly-broken power law. This finding is analogous to previous measurements of the stellar halo. Figure \ref{fig:5d mock} shows the posterior distribution of such a fit. We see that the model has been strongly constrained to an oblate spheroid ($q=0.84$) with a single breaking radius in between the two true breaking radii. This result is a cautionary tale: despite the posterior being strongly constrained to be an oblate spheroid with a single break, the underlying true density can be triaxial and tilted. Thus, there is strong motivation to project the best-fit models back onto data space, where clear systematics (such as misinterpreting a triaxial tilted ellipsoid as an oblate spheroid) would be visible as residuals between the data and the model. This lesson forms the basis of how we examine and present our results in the following section.
\begin{figure*}
    \centering
    \includegraphics[width=\textwidth]{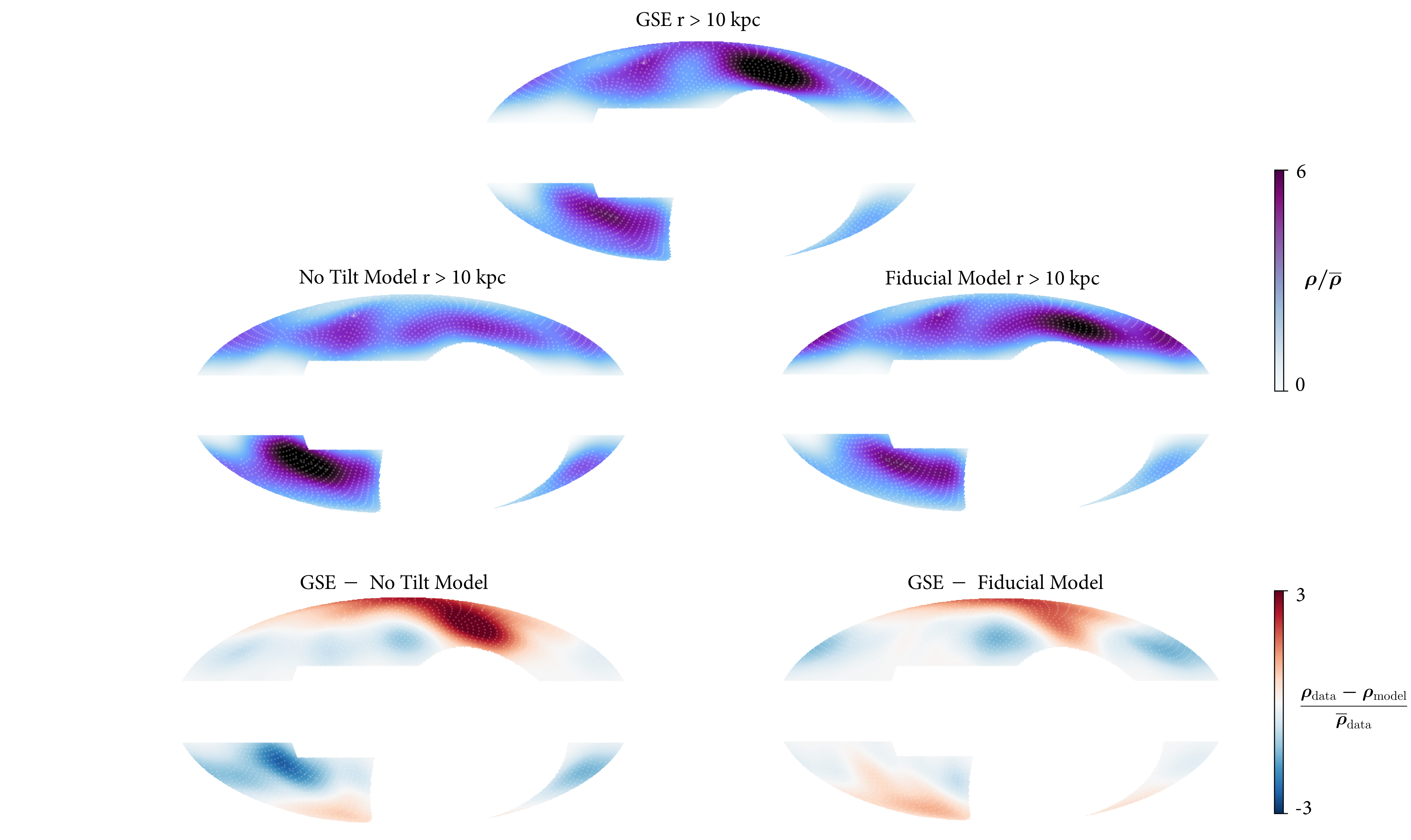}
    \caption{On-sky stellar density of the data and of the models projected onto data space. The top panel shows the density of GSE, and middle left (right) shows the density of a non-tilted (fiducial) model. Densities are normalized by the mean density in each panel. The bottom left (right) panel shows the residuals of the data subtracted by the non-tilted (fiducial) model, normalized by the mean density of the data. The fiducial model is a tilted, triaxial ellipsoid, and the non-tilted model is created by fixing all other parameters of the fiducial model and setting the rotation angles (both pitch and yaw) to zero. All panels are smoothed with a $20^\circ$ Gaussian kernel. The faint white dots mark H3 tiles. The top and middle panels include the significant effect of the H3 selection function. However, the model residuals in the bottom panels are insensitive to the selection function.}
    \label{fig:onsky}
\end{figure*}

\begin{figure*}
    \centering
    \includegraphics[width=0.8\textwidth]{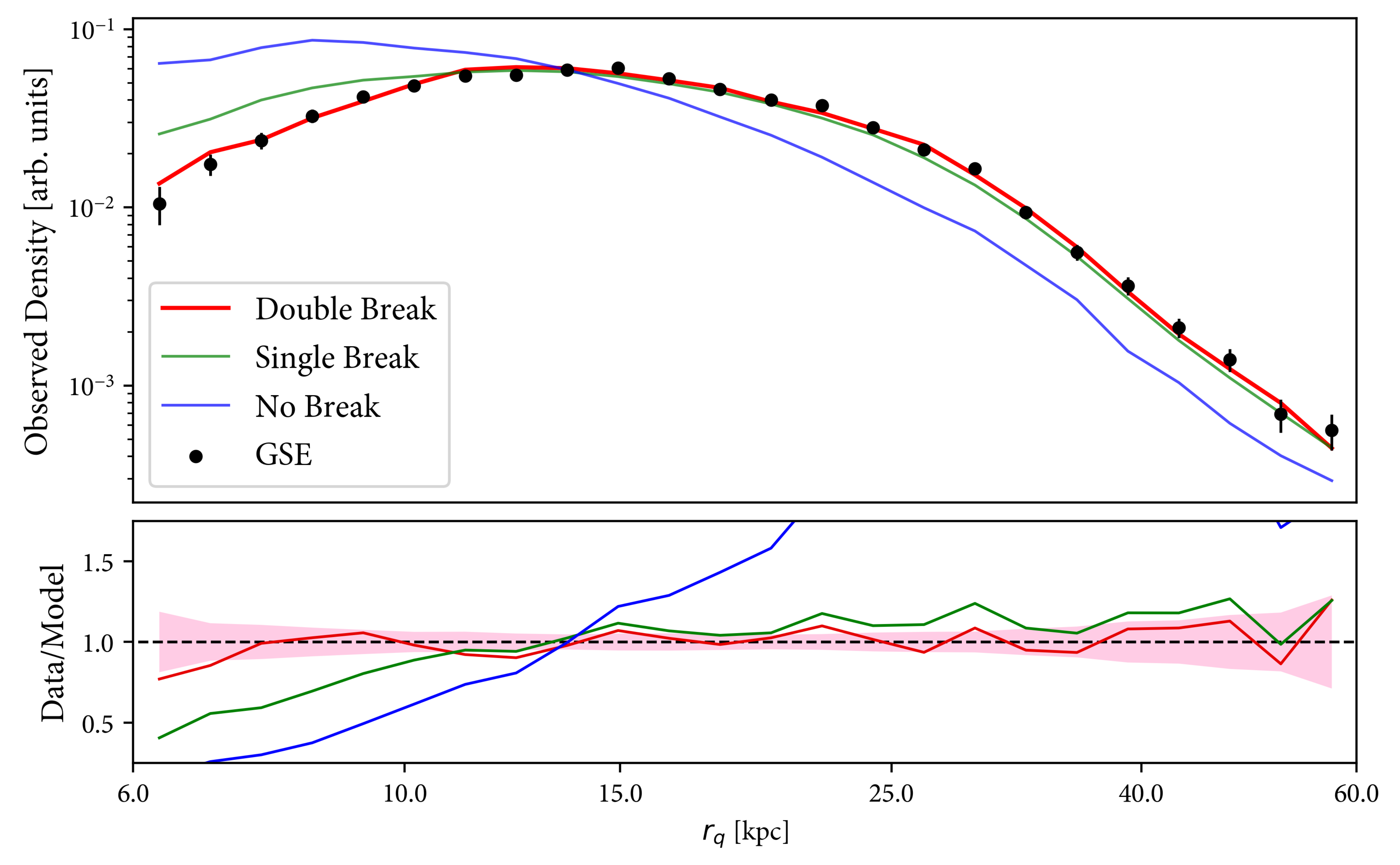}
    \caption{Observed radial density distribution of GSE (black symbols), compared to best-fit power laws with varying numbers of breaking radii (solid lines). The best-fit models have been convolved with the H3 selection function.  The Poisson uncertainty in the data (shaded in red in the bottom panel) is derived by producing 100 mock catalogs and computing the variance. The single break power law fits the outer halo well, but fails to simultaneously model the inner $r_q<15\text{ kpc}$ halo. The double break power law fits the data across the whole radial range within the model uncertainties. Note that the apparent decline at $r_q<15\text{ kpc}$ is due to selection effects. The true density distribution is a monotonically decreasing function of radius.}
    \label{fig:rqres}
\end{figure*}

\section{Results} \label{sec:results}

Here we present the results of measuring the shape and density profile of GSE. We first present results based on our fiducial model. We decompose the density profile into its angular and radial components, and compare the best-fit model and the data in each projection. Finally, we discuss variations to the fiducial model and sample, and how that affects our fits.

\subsection{Fiducial Model:\\Tilted \& Doubly Broken}

Our fiducial model for the stellar halo is a triaxial, doubly broken power-law with an orientation not constrained to lie in the Galactic plane. There are three key features to our best-fit model: 1) we find two breaking radii at $12 \text{ kpc}$ and $28\text{ kpc}$; 2) the principal axes ratios are $1:p:q=10:8:7$, which is nearly prolate; 3) we find that the ellipsoid is tilted by $25^\circ$ off of the Galactic plane and $24^\circ$ away from the Sun-Galactic Center axis.   Figure \ref{fig:10d corner} shows the full posterior of each parameter of the fiducial model. That the covariance among parameters is mostly small indicates that our model parameters are not degenerate with one another. This is consistent with our expectation that each parameter should affect distinct spatial features of the ellipsoid. In Figure \ref{fig:fiducial model} we plot the isodensity surface of the fiducial model in Galactocentric coordinates, evaluated at $r_q=20\text{ kpc}$. The solar location is marked as a red star, and an artist impression of the Galactic disk \citep{MWdisk} is overplotted to orient the reader.

In figure \ref{fig:onsky} we show the on-sky stellar density of the data and of the models projected onto data space. The top panel shows the density of GSE, and middle left (right) shows the density of a non-tilted (fiducial) model. Densities are normalized by the mean density in each panel. The bottom left (right) panel shows the residuals of the data subtracted by the non-tilted (fiducial) model, normalized by the mean density of the data. The fiducial model is a tilted, triaxial ellipsoid, and the non-tilted model is created by fixing all other parameters of the fiducial model and setting the rotation angles (both pitch and yaw) to zero. All panels are smoothed with a $20^\circ$ Gaussian kernel. The faint white dots mark H3 tiles. The top and middle panels reflect the H3 selection function to first order. However, since the data and the models are subject to the same selection function, the difference in bottom panel model residuals are insensitive to the selection function. In this context, we observe a striking dipolar over-underdensity pair in $(-l,+b)$ and $(+l,-b)$ for the non-tilted model residual plot. This dipole is bookended by the Virgo Overdensity in the North and the Hercules-Aquila Cloud in the South. The dipole residual is greatly reduced in the tilted (fiducial) model, demonstrating that an overall rotation to the ellipsoid can explain the observed dipole in the non-tilted residual. However, the residuals of the fiducial model is still imperfect, which we discuss in Section \ref{sec:discussion}.

In Figure \ref{fig:rqres} top panel we plot the observed radial density of GSE (black dots) and compare to the fiducial model (red line) as well as best-fit singly-broken (green line) and no-break (blue line) power law models. In the bottom panel we plot the residuals of each model to the data, and indicate the Poisson uncertainty of the data in red shading, evaluated from 100 mock catalogs. While the single break power law fits the $r_q>15\text{ kpc}$ halo well, it over predicts the stellar density at inner radii. To model both the inner and outer halo, the data requires a minimum of two breaking radii. Using the Akaike information criterion \citep[AIC;][]{AIC}, we can use the likelihood values to determine which model better describes the data. AIC is defined as $2k - 2 \ln \mathcal{L}$, where $k$ is the number of free parameters and $\mathcal{L}$ is the maximum likelihood value for the model. Based on this criterion, the probability that the singly-broken power law minimizes the information loss compared to the doubly-broken power law is $e^{-23724}$, essentially zero. We thus confidently rule out the singly-broken power law as the preferred model.

It is worth nothing that the $r<10\text{ kpc}$ halo strongly overlaps with in-situ stars (both the disk and in-situ halo), and many prior studies of the stellar halo therefore exclude this region from their sample. Indeed, the inner breaking radius would be difficult to observe if we were not able to separate GSE from the in-situ stars. Owing to the clean chemistry and kinematic selection enabled by H3, we are able to probe the density profile to relatively small Galactocentric radii. Furthermore, if our sample were significantly contaminated by disk stars, the innermost power law slope would likely be steeper and the difference between $\alpha_1$ and $\alpha_2$ would be smaller than what we measure. Hence, our measurement of the innermost breaking radius is likely not a product of disk contamination. Lastly, $\alpha_1$ and $\alpha_2$ are $\sim8 \sigma$ apart from each other and $\alpha_2$ and $\alpha_3$ are $\sim15 \sigma$ apart, demonstrating the statistical significance of the doubly-broken power law.

\begin{figure}
    \centering
    \includegraphics[width=.5\textwidth]{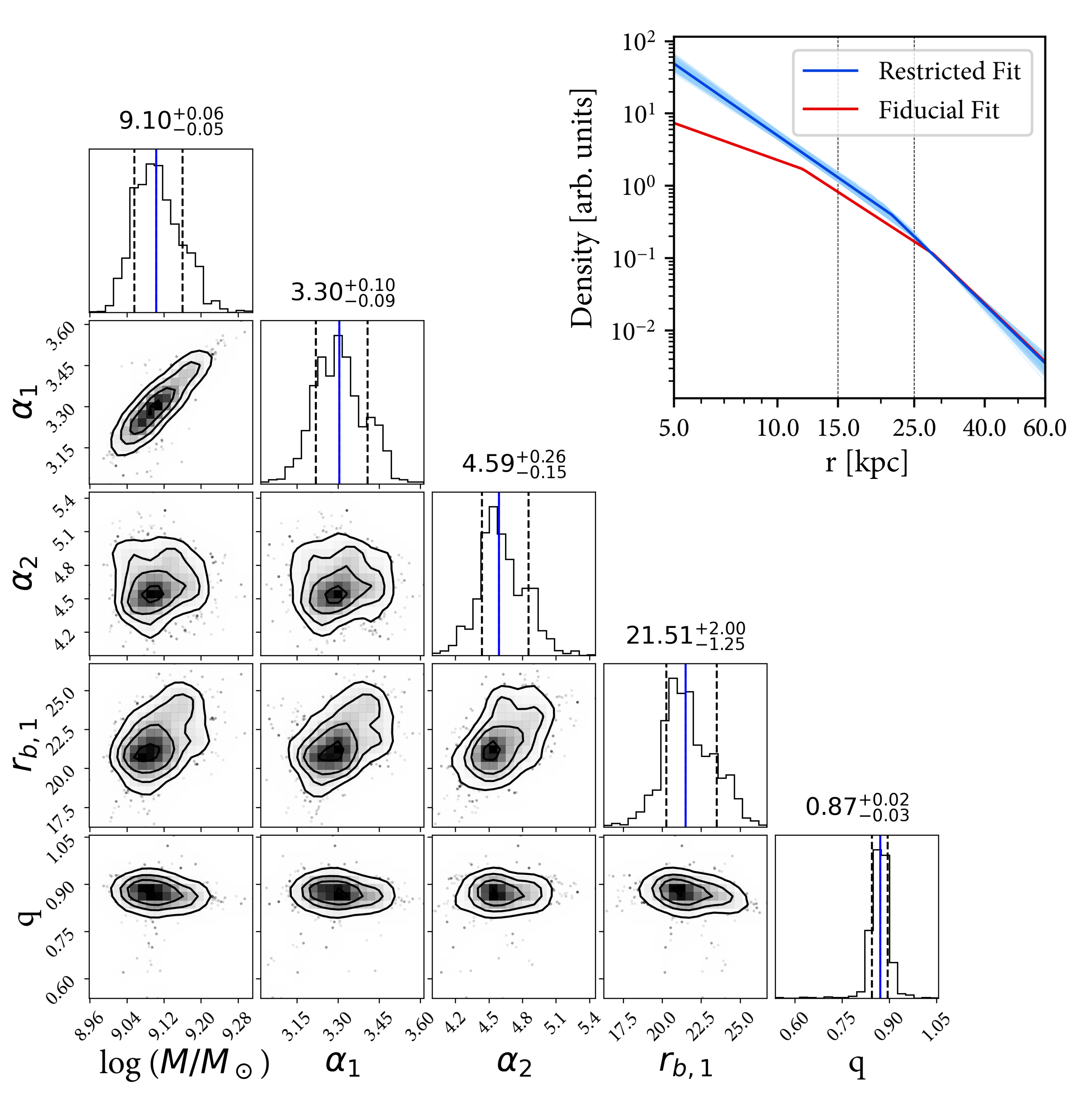}
    \caption{Posterior distribution from fitting a halo sample selected purely on $|Z|>4\text{ kpc}$. We restrict the model to be aligned with the Galactic disk, and allow for only one breaking radius. The resulting parameters we derive are comparable to literature values. Blue lines indicate the mean, and dotted lines mark the $16^{th}$ and $84^{th}$ percentiles. The radial density profile resulting from the restricted fit is plotted in the upper right corner in blue, and the fiducial fit is drawn in red for comparison. The radial densities are in arbitrary units.}
    \label{fig:5dxue}
\end{figure}

\subsection{Variations to the Fiducial Model and Sample}

In this paper we have explored both a new model (triaxial ellipsoid that is allowed to rotate) and new data (a high-purity GSE sample obtained from chemodynamical selection) to measure the density profile of the most dominant contributor to the stellar halo. It is important to understand how our results depend on each component.

From Figure \ref{fig:5d mock}, we have learned that an intrinsically triaxial and tilted ellipsoid can be incorrectly constrained to be an oblate spheroid if the model is restricted to be aligned with the Galactic plane. Thus, if we fit our fiducial GSE sample with a model that is restricted to be aligned with the plane, we would obtain an oblate spheroid as our best-fit model. However, we know from Figure \ref{fig:onsky} and Figure \ref{fig:rqres} that this oblate spheroid model does not reproduce either the radial or angular distribution of the data. Thus, this exercise once again reveals the limitations of fitting a restricted model to the data, and the value of a generative framework that allows one to evaluate the model in data space.

To connect our data to previous studies, we fit a restricted model to a sample (an oblate spheroid with a singly-broken power law) to a halo sample selected only on $|Z|>4\text{ kpc}$, with no no chemical or kinematic selection. Such a geometric selection of halo stars being above some Galactic height is characteristic of pre-\textit{Gaia} halo samples \citep[e.g.,][]{xue15}. Our sample is of course still specific to the H3 footprint, so it is not identical to prior studies. Figure \ref{fig:5dxue} shows the posterior of this restricted fit. Similar to previous work, we find that the data can be strongly constrained to an oblate spheroid ($q\sim0.9$) with one breaking radius at $\sim20\text{ kpc}$. The inner power law slope, $\alpha_1=3.3$ is on the steeper end of literature values, and the outer power law slope, $\alpha_2=4.6$, is near the average of literature values. We further elaborate on placing our results in the context of the literature in the Discussion.

We now assess the specific choice of eccentricity used to define GSE. To this end, we return to Figure \ref{fig:espace}, where we show the distribution of eccentricity against metallicity of the parent sample. While GSE dominates the high-$e$ sample, it does extend down to lower eccentricities. It is thus worthwhile to explore how our result changes if we relax the eccentricity selection of GSE to be $e>0.5$. This increases our GSE sample size to 7886. The result is summarized in Table \ref{table:summary}. Other than the total stellar mass, which increases proportionally to the sample size as expected, none of the density parameters change in any appreciable way. This also demonstrates that our kinematic cut does not induce large selection effects in the radial range that we explore in this study.

Finally, we explore how our result changes in the complete absence of a chemodynamical selection by fitting the whole parent sample with the fiducial model. This sample includes the thick disk and in-situ halo, and reflects the total sum of distant ($d>1\text{ kpc}$) giants in H3 that are not part of a kinematically cold structure (i.e., Sagittarius stream, Sextans, Ursa Minor, Palomar 13, Draco, M31, M33, M92, and GD-1). The result is summarized in Table \ref{table:summary}. The tilt angle, while still nonzero, is notably smaller ($\sim10^\circ$) and the shape is closer to oblate (10:9:8), which is what we expect since the thick disk and in-situ halo occupy a distribution that is different from GSE and is aligned with the Galactic plane. Additionally, the inner power-law slope is steeper ($\alpha_1\sim2.1$) than the fiducial model, again likely due to the thick disk and in-situ halo as their distributions are steeper than GSE and occupy the inner 10 kpc.

\begin{figure*}
    \centering
    \includegraphics[width=0.85\textwidth]{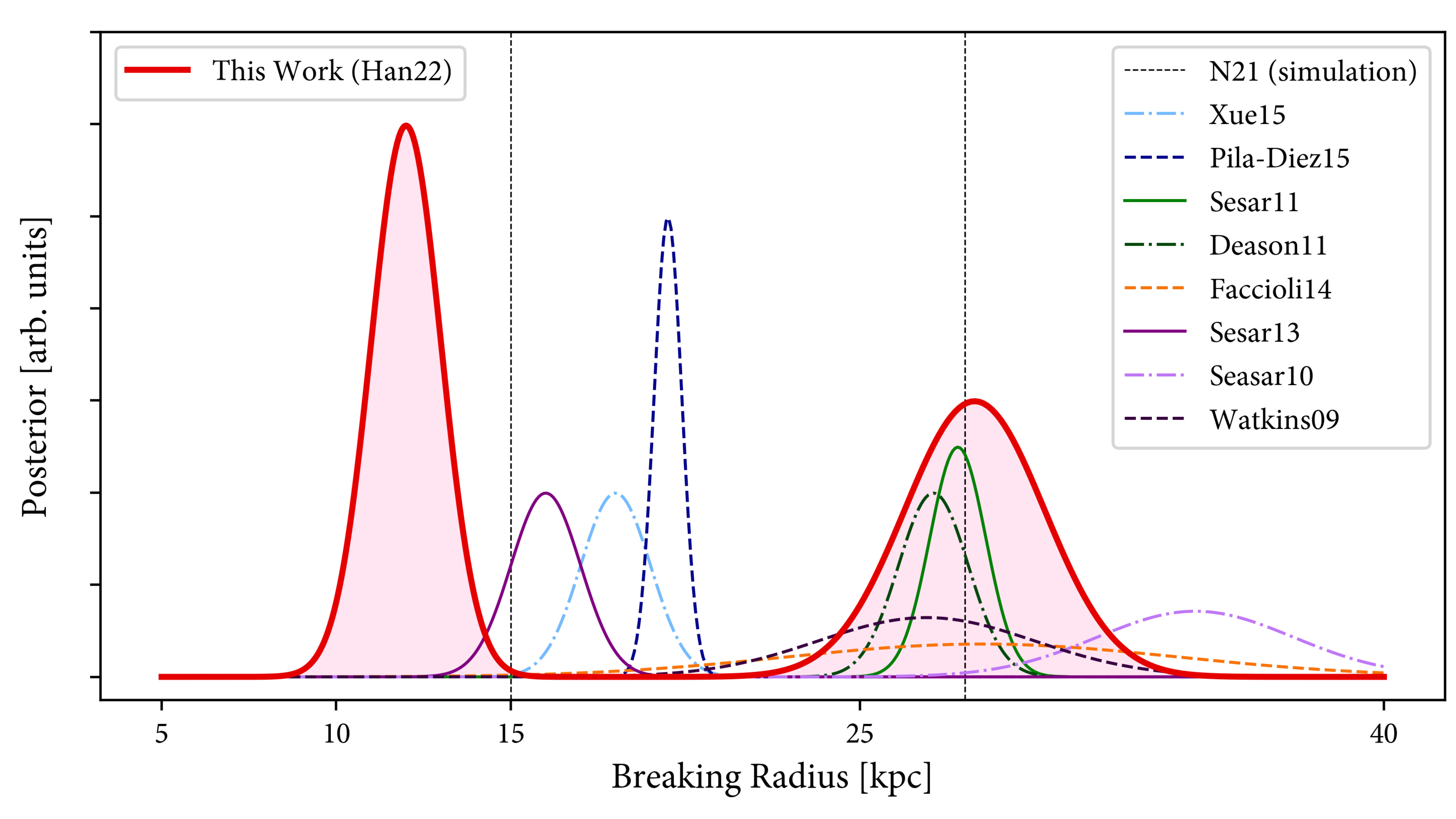}
    \caption{Posterior probability distribution of  breaking radii, comparing our best-fit double-break (red shaded regions) to previous work.  This plot demonstrates the dichotomy in the literature values of the breaking radius, clustering at either $\sim20\text{ kpc}$ or $\sim30\text{ kpc}$. This work simultaneously measures breaking radii in the inner and outer halo.}
    \label{fig:breaking radii}
\end{figure*}

\begin{figure}
     \centering
     \includegraphics[width=.48\textwidth]{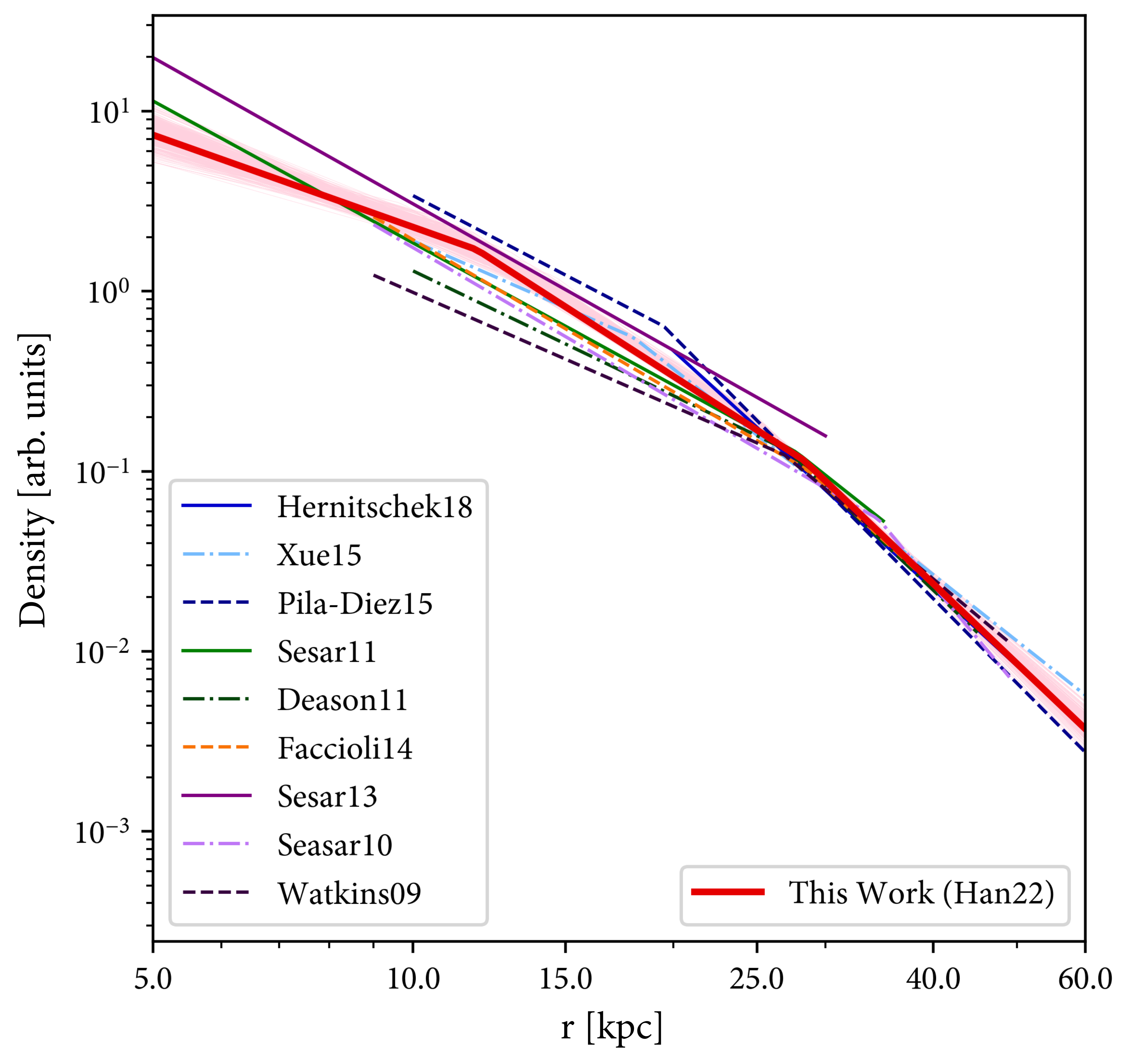}
    \caption{Radial density distributions of the stellar halo, comparing the result from this work to the literature. Densities are plotted over the radial range within which they were measured. Our double-break model is plotted in red, along with 500 draws from the posterior in thinner red lines to demarcate the statistical uncertainties. This model roughly traces the average literature profile in the inner ($r<15 \text{ kpc}$), mid ($15 \text{ kpc}<r<25 \text{ kpc}$), and outer ($r>25 \text{ kpc}$) halo, thus resolving the tension in density profiles measured for the inner and outer halo.}
    \label{fig:theory}
\end{figure}

\section{Discussion} \label{sec:discussion}

\subsection{Caveats and Limitations}

In this study, we have explored a triaxial ellipsoid model to measure the shape and density profile of the most dominant component of the accreted stellar halo of the Galaxy. This model reproduces key features of the data in both radial and angular projections, corresponding to the overall tilt of the halo and the double break in the power law. We thus believe the fiducial model to be representative of the underlying data. However, the true stellar halo is surely not a perfect ellipsoid as we have drawn in Figure \ref{fig:fiducial model}. While the on-sky density residuals of the fiducial model is clearly better than that of the non-tilted model in Figure \ref{fig:onsky}, we still observe a spatially correlated offset from the data. This is likely due to a more complex shape of the halo, such as its overall orientation changing as a function of radius \citep[e.g.,][show this in simulations]{prada19, emami21, shao21} and/or having different radial density distributions on either side of the Galactic center due to dynamical friction as the progenitor sinks into the Galaxy \citep[][]{n21}. While our work allows for a more general shape than in previous work, there are clear motivations to consider even more flexible models. Factoring in the in-situ disk and halo, the combined halo is likely a superposition of at least two distinct shapes: a triaxial ellipsoid accreted halo, and an oblate in-situ halo.

Another limitation of this study is that H3 does not contain data below $\delta<-20^\circ$, so we cannot measure features to the halo profile contained in those regions. Future spectroscopic surveys in the Southern hemisphere such as 4MOST and SDSS-V will provide important constraints on the behavior of the stellar halo. Furthermore, the distances estimated by \texttt{MINESWeeper} have $\sim10\%$ statistical uncertainties that we do not directly factor into the inference framework. Instead, we verify that these uncertainties do not significantly affect our results by fitting mock catalogs that include distance uncertainties. All parameters are inferred to be well within statistical uncertainty from their true values, as we show in Figure \ref{fig:mock corner}. We have also removed very distant stars beyond $r>60\text{ kpc}$ that are most affected by distance uncertainties and have poor astrometric quality. To properly model the outer halo, where tracers are rare and distance uncertainties can sway the results, one should marginalize the likelihood function over the distance posterior. The constituents of the distant $r>60\text{ kpc}$ halo is largely uncharted, and will be an important direction for future work. Our work also does not attempt to model the spatial-dependent kinematics of GSE which could introduce systematic errors (e.g., orbits becoming more circular in the outer halo, where unrelaxed tidal debris may still exist). Lastly, while the fiducial $e>0.7$ sample should largely be comprised of GSE, the $e>0.5$ sample may contain other mildly eccentric, accreted substructures such as Sequoia \citep{Myeong19, matsuno19}, Arjuna, and I'itoi \citep{naidu20}. However, these structures are subdominant (Sequoia, for example, is estimated to be $\sim40$ times less massive than GSE). Thus, their contribution to the density profile is very small compared to GSE. Nonetheless, understanding the spatial distribution of these structures is important to completing our knowledge of the accreted stellar halo.

\subsection{Comparison to Previous Work}

Due to our novel sample comprising high-probability GSE stars, it is not a straightforward task to directly compare our results to previous studies. Instead, here we place our fiducial density model in the context of previous studies of the stellar halo.

We first highlight the double-break feature of our density profile. Figure \ref{fig:breaking radii} shows the distribution of breaking radii $r_b$ in the literature compared to this work. Previous results report a single breaking radius, and we visualize these radii as Gaussian distributions based on the reported uncertainties. This figure highlights the dichotomy in the literature values of $r_b$: there is a cluster of $r_b\sim 15-20\text{ kpc}$ and a cluster of $r_b\sim 25-30\text{ kpc}$. If we interpret the double-break power law that we measure to be the outer and inner apocentric pile-ups of GSE, we can understand the dichotomy as an artifact of the survey footprint and radial ranged probed by previous studies. Depending on the specific sightline, the outer (inner) apocentric pile-up can be more prominent than the inner (outer) pile-up due to the tilt of GSE. Additionally, because GSE is triaxial, the physical location of the breaking radius can change up to the major to minor axis ratio depending on the sightline, which is a $30\%$ difference in our model. Our study accounts for all of these effects, and we find that our two measured breaking radii coincides with the two clusters of results in the literature.

\begin{figure*}
    \centering
    \includegraphics[width=0.75\textwidth]{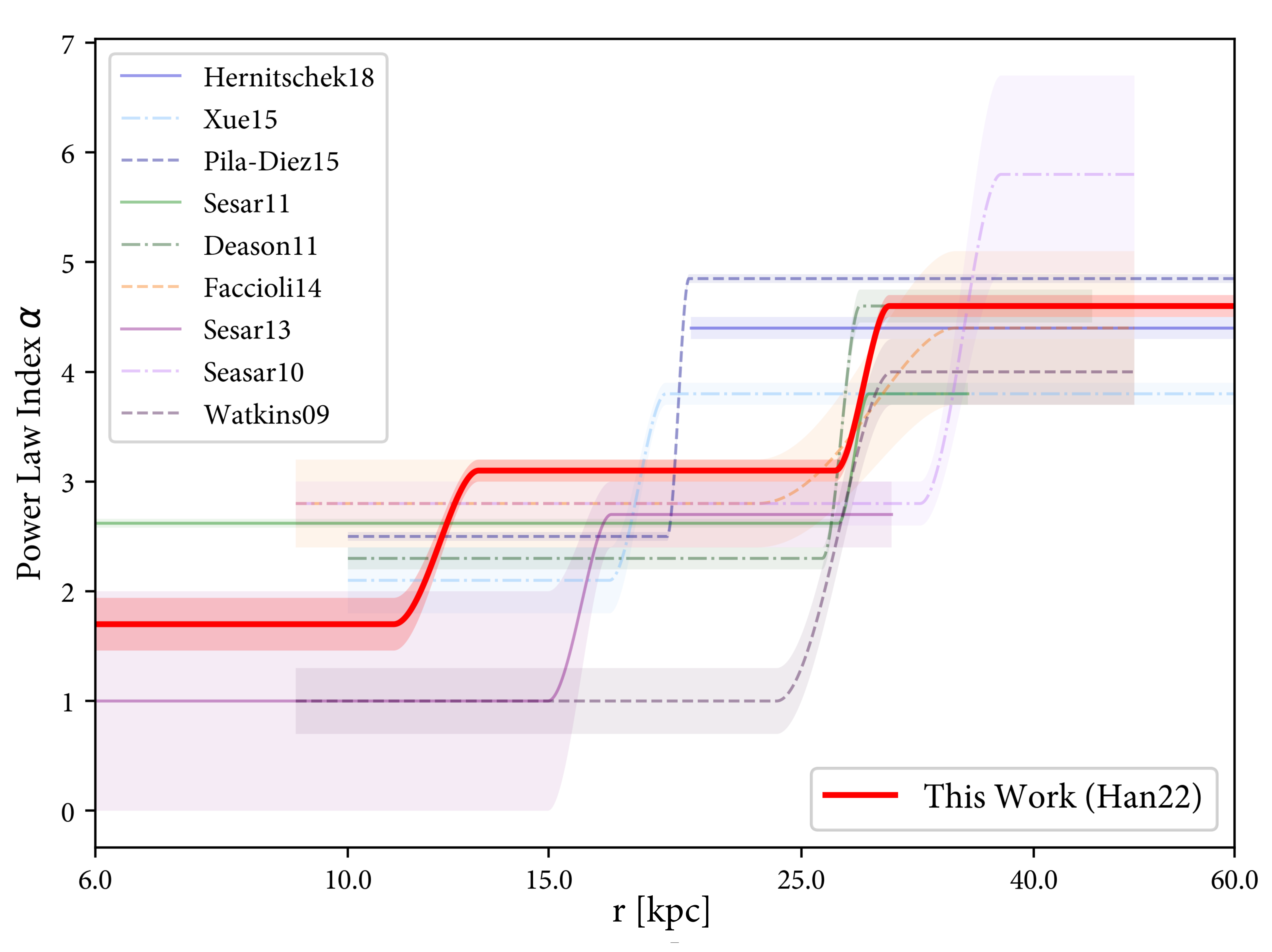}
    \caption{Power law indices as function of Galactic radius, comparing the result of this work (red line) to the literature. Each result is represented as a line with shaded regions demarcating the statistical uncertainty. Along each line, the transition region between power law indices corresponds to the uncertainty in breaking radius.}
    \label{fig:tron}
\end{figure*}

In Figure \ref{fig:theory} we plot the density profiles from the literature over their respective radial range of the study. In Figure \ref{fig:tron} we make a similar comparison for the power law index, where the transition region between inner and outer power law indices correspond to reported uncertainties of the breaking radius. We see that the density profile presented in this study approximately traces the average literature profile in the inner ($r<15 \text{ kpc}$), mid ($15 \text{ kpc}<r<25 \text{ kpc}$), and outer ($r>25 \text{ kpc}$) halo. 

A noteworthy feature is that our inner breaking radius is smaller than literature estimates. Several factors could lead to this result. First, our halo sample includes stars at smaller Galactocentric radius than most studies owing to our ability to effectively remove in-situ stars. Second, from the mock test shown in Figure \ref{fig:5d mock} we have seen that by approximating a singly-broken power law to an intrinsically doubly-broken power law, we estimate a breaking radius that is greater than the inner breaking radius. From this we can infer that even a study that surveys sightlines dominated by the inner breaking radius, stars from the outer breaking radius will shift the measured breaking radius outwards. On the contrary,  previous samples dominated by stars at greater distances are not sensitive to the inner breaking radius. 

We now discuss the shape and orientation of our model. The vast majority of previous studies report an oblate spheroid that is aligned with the disk, which has principal axes ratios of 1:1:$0.6-0.8$ \citep[][and references therein]{gerhard-BH16}. Among the few studies that have fit a triaxial ellipsoid, \citet{iorio18} report axes ratios of 1:0.8:$0.5-0.6$ from \textit{Gaia} DR1 RRL cross-matched with 2MASS out to $r_{\text{Gal}}<28\text{ kpc}$. This axis ratio is comparable to what we find, which is 1:0.8:0.7 as shown in Figure \ref{fig:fiducial model}. In terms of orientation, most studies assume a fixed orientation of the halo that is aligned with the disk. However, \citet{IB19} report a misalignment (yaw) angle of the major axis with the Galactic X axis by $70^\circ$ based on DR2 RRL. The yaw angle reported by \citet{IB19} and our study both point towards the same quadrants in the Galactic XY plane, although the precise value of the angle is different ($24^\circ$ from the X axis in our study). Given that the sample used in this study is significantly different from the \textit{Gaia} RRL due to our chemodynamical selection, this difference in yaw angle may not be significant.  Future work should clarify this issue. Furthermore, \citet[][]{IB19} speculate the existence of a tilt of the halo with respect to the Galactic plane based on the XZ projection of RRL, and the direction of the tilt is consistent with what we report in this paper. In both studies, the ellipsoid is tilted above the plane in the direction of the Sun.

Finally, we discuss our estimate of the total stellar mass of GSE. As summarized in Table \ref{table:summary}, we infer a total GSE stellar mass of $5.8 - 7.6 \times 10^8 M_{\odot}$, depending on if we define GSE to be $e>0.7$ or $e>0.5$. As we do not have data within $r_{\text{Gal}}<5\text{ kpc}$, the mass enclosed in this region is extrapolated from the inferred density profile, accounting for $14\%$ of the total mass. The inferred mass range of GSE comfortably brackets the literature values of the GSE mass such as \citet{Helmi18} $6\times 10^8 M_{\odot}$, \citet{vincenzo19} $5\times 10^8 M_{\odot}$, \citet{feuillet20} $7\times 10^8 M_{\odot}$, and \citet{naidu20} $4-7\times 10^8 M_{\odot}$. There exist exceptions, such as \citet[][]{mackereth18} who find a lower accreted stellar halo mass of $3\times 10^8 M_{\odot}$ at $e>0.7$. We note that the restricted fit we present in Figure \ref{fig:5dxue} yields a significantly higher mass estimate of $1.3\times 10^9 M_{\odot}$ than the fiducial fit, which shows that the triaxiality and the existence of an inner break (thus allowing for a shallower innermost power law slope) can significantly affect the total stellar mass estimate. This higher mass estimate is remarkably similar to what \citet{deason19} find for the total stellar halo mass based on RGB stellar counts, $1.4\times 10^9 M_{\odot}$. 

\subsection{Implications}

A tilted, doubly-broken density profile of the stellar halo has a number of implications for the Galaxy. First, the overall tilt of the stellar halo with respect to the Galactic disk serves as independent evidence for the accreted origin of the halo, as it is difficult to achieve such a configuration through purely in-situ processes. The two density breaks corroborate the scenario in which GSE goes through two apocentric passages before merging with the Galaxy, as predicted by \citet{n21}. The second apocenter occurs at smaller Galactocentric radius than the first due to dynamical friction, and the distance between them is an observable quantity that we have measured in this work that can be used to constrain properties of the merger. For example, the total mass ratio and the concentration of the progenitor should strongly affect how many apocentric passages the progenitor can go through before fully merging, and how deeply the progenitor plunges at each passage. Future work will address such questions. 

Second, the association of the tilted stellar halo with an ancient ($t_{lb}\sim10\text{ Gyr}$) merger holds clue to the shape of the dark matter halo. \citet{han22} show that a present-day spatial asymmetry of GSE, coupled with its estimated old age, implies that at least some fraction of the underlying dark matter halo is also tilted (hence necessarily aspherical) and oriented in a similar direction as GSE. Understanding the dark matter distribution in the Galaxy is important for both Galactic archeology and local direct detection experiments, as the nature of the DM halo affects the local DM density and velocity distributions \citep[e.g.,][]{evans19,necib19}. Furthermore, while the evolution of the stellar disk induced by the dark matter halo has been studied before \citep[e.g.,][]{debattista13,yurin15,dillamore21}, the stability and growth of the disk specifically in a tilted, triaxial dark matter halo is an open question, and will be investigated in future studies.

Lastly, it has been recognized that the Large Magellanic Cloud (LMC) likely has a significant influence on the global shape of the halo \citep[e.g.,][]{GC19, erkal20, conroy21, PP21}. While this effect is most prominent at larger distances ($r_{\text{Gal}}>50\text{ kpc}$) than what we consider here (see \citealt{lucchini21} for an alternative scenario), the direction of overdensities due to the LMC and GSE have similar on-sky orientations. Whether this curious alignment is due to coincidence or a manifestation of the large-scale environment that our Galaxy resides in will be addressed by future studies.

\section{Summary}

In this paper we have used the H3 Survey to infer the shape and density profile of GSE, the most dominant component of the accreted stellar halo. We employed both chemical and kinematic selection to obtain a high-purity sample of halo stars belonging to GSE, to which we fit a general ellipsoid with two rotation angles and a multiply-broken power law along its flattened radius. Table \ref{table:summary} outlines the best-fit parameter values for the various combination of models and samples that we describe in Section \ref{sec:results}. The key findings from this work can be summarized as follows.

\begin{enumerate}

    \item The shape of GSE can be well-fit by a tilted, triaxial ellipsoid. The principal axes ratios are 10:8:7, and the major axis of the ellipsoid points $25^\circ$ above the Galactic plane towards the Sun. The overall tilt of the halo is evident in the data, shown in Figure \ref{fig:onsky}.
    
    \item The radial density profile of GSE is well described by a doubly-broken power law, as shown in Figure \ref{fig:rqres}. The two breaking radii are at 12 kpc and 28 kpc. We interpret these breaking radii as representing two apocenters of the decaying orbit of GSE before it fully merged with the Galaxy. We rule out a singly-broken power law with high confidence.
    
    \item Integrating the density profile over the entire halo, we find the stellar mass of GSE is $5.8 - 7.6 \times 10^8 M_{\odot}$, depending on the precise eccentricity selection used to define GSE.
    
\end{enumerate}

Ongoing and upcoming surveys such as DESI, SDSS-V, WEAVE, and 4MOST will collect considerably more data in the coming years. These data will enable stronger constraints on the shape and orientation of the stellar halo.  Future directions include probing the halo beyond 60 kpc and constraining variation of its shape and orientation as a function of radius.  Such measurements will provide even sharper constraints on the assembly history of our Galaxy.

\begin{deluxetable*}{cccccccccccc}[ht]
\tabletypesize{\footnotesize}
\tablecolumns{11}
\tablewidth{0pt}
\tablecaption{ Summary of Best-Fit Parameters \label{table:summary}}
\tablehead{
\colhead{Sample} &
\colhead{$\log (M/M_\odot)$} &
\colhead{$\alpha_1$} &
\colhead{$\alpha_2$} & 
\colhead{$\alpha_3$}&  
\colhead{$r_{b,1}$ [kpc]} &
\colhead{$r_{b,2}$ [kpc]} & 
\colhead{p} &
\colhead{q} & 
\colhead{pitch $(^\circ)$} & 
\colhead{yaw $(^\circ)$}}
\startdata
\\
\multicolumn{11}{c}{Fiducial Model}\\
\hline\\\vspace{0.2cm}
Fiducial GSE& ${8.76}^{+0.02}_{-0.02}$ & ${1.70}^{+0.16}_{-0.24}$ & ${3.09}^{+0.10}_{-0.11}$ & ${4.58}^{+0.10}_{-0.11}$ & ${11.85}^{+0.92}_{-0.71}$ & ${28.33}^{+1.05}_{-1.55}$ & ${0.81}^{+0.03}_{-0.03}$ & ${0.73}^{+0.02}_{-0.02}$ & ${-25.39}^{+3.11}_{-3.20}$ & ${-24.33}^{+4.94}_{-5.51}$ &
\\
\vspace{0.2cm}
$e>0.5$ GSE &
${8.88}^{+0.02}_{-0.02}$ & ${1.70}^{+0.05}_{-0.06}$ & ${3.10}^{+0.06}_{-0.06}$ & ${4.65}^{+0.03}_{-0.04}$ & ${10.61}^{+0.43}_{-0.46}$ & ${29.26}^{+0.90}_{-1.07}$ & ${0.82}^{+0.02}_{-0.02}$ & ${0.74}^{+0.02}_{-0.02}$ & ${-25.13}^{+2.02}_{-1.97}$ & ${-23.88}^{+3.36}_{-3.36}$ &
\\
\vspace{0.2cm}
Parent&
${9.13}^{+0.04}_{-0.03}$ & ${2.08}^{+0.24}_{-0.20}$ & ${3.45}^{+0.10}_{-0.09}$ & ${4.45}^{+0.15}_{-0.16}$ & ${9.48}^{+0.86}_{-0.91}$ & ${27.68}^{+1.92}_{-1.67}$ & ${0.88}^{+0.02}_{-0.02}$ & ${0.78}^{+0.01}_{-0.01}$ & ${-10.40}^{+2.08}_{2.09}$ & ${-11.06}^{+4.53}_{-4.15}$ & 
\\ \\
\hline
\\
\multicolumn{11}{c}{Restricted Model: single-break, p=1, pitch=0, yaw=0} \\
\hline
\\
$|Z|>4\text{ kpc}$& ${9.10}^{+0.06}_{-0.05}$ & ${3.30}^{+0.10}_{-0.09}$ & ${4.59}^{+0.26}_{-0.15}$ & - & ${21.51}^{+2.00}_{-1.25}$ & - & 1 (fixed) & ${0.87}^{+0.02}_{-0.03}$ & 0 (fixed) & 0 (fixed) &
\\
\vspace{0.1cm}
\enddata
\end{deluxetable*}

\clearpage

\bibliography{bibi}{}
\bibliographystyle{aasjournal}

\appendix
In Table \ref{table:selfunc} we outline the target selection criteria for H3 (more details will be presented in a future data release). The \texttt{mgiant} selection is based on WISE colors as described in \citet[][in spite of the name, the selection is mostly sensitive to luminous K giants]{conroy18}. The \texttt{bhb} selection is based on color cuts presented in \citep[][]{deason12}, and the \texttt{rrl} sample is taken directly from \cite{sesar17}. We note that there also exist rank 3 and rank 4 targets; however, they comprise less than 1\% of the relevant sample, so we exclude those targets in this study. Color-based tiles (ca1, cb1, cc1) were used pre-\textit{Gaia}, and the only change among the c-tiles are subtle changes to the definition of rank-1 stars ($\sim6\%$ of the data), while rank-2 stars ($\sim93\%$ of the data) are defined the same. The \textit{Gaia}-based tiles (ga1, gb1, gc1, gf1) use \textit{Gaia} parallax as the main selection. Between ga1 and gb1 the only change is in the magnitude range of rank-1 stars. From gb1 to gc1 and gf1, the parallax limit is changing to more efficiently target stars that are further away and likely in the halo. From gc1 to gf1, we update the parallax selection based on \textit{Gaia} EDR3 \citep{EDR3} which has more precise astrometry. Note that we skip ``gd1" and ``ge1" to avoid confusion with the GD-1 stellar stream \citep{GD-1} and \textit{Gaia}-Sausage-Enceladus \citep{belokurov18,Helmi18}. The typical number of giants in each tile is $\sim5-15$.

\begin{deluxetable*}{cccc}[h!]
\tabletypesize{\footnotesize}
\tablecolumns{4}
\tablecaption{ H3 Target Selection \label{table:selfunc}}
\tablehead{
\colhead{selection ID} &
\colhead{rank 1} &
\colhead{rank 2} & \colhead{\# of Tiles (\% of Data)}}
\startdata
\\
ca1 & \texttt{mgiant} and  ($g-r<1$) and ($14<r$)&
(\texttt{not} rank1) and ($g-r<1$) and ($15<r<18$) & 27 (2\%)
\\\\
cb1 & (\texttt{mgiant} or \texttt{bhb}) and ($14<r$) &
(\texttt{not} rank1) and ($g-r<1$) and ($15<r<18$) & 2 (0.2\%)
\\\\
cc1 & (\texttt{mgiant} or \texttt{bhb} or \texttt{rrl}) and ($14<r$) &
(\texttt{not} rank1) and ($g-r<1$) and ($15<r<18$) & 75 (5\%)
\\\\
ga1 & (\texttt{mgiant} or \texttt{bhb} or \texttt{rrl}) and ($14<r<18$) and ($\pi<0.5$)&
(\texttt{not} rank1) and ($\pi - 2\sigma_\pi<0.5$) and ($15<r<18$) & 158 (12\%)
\\\\
gb1 & (\texttt{mgiant} or \texttt{bhb} or \texttt{rrl}) and ($13.5<r<17.5$) and ($\pi<0.5$)&
(\texttt{not} rank1) and ($\pi - 2\sigma_\pi<0.5$) and ($15<r<18$) & 95 (8\%)
\\\\
gc1 & (\texttt{mgiant} or \texttt{bhb} or \texttt{rrl}) and ($13.5<r<17.5$) and ($\pi<0.3$)&
(\texttt{not} rank1) and ($\pi <0.4$) and ($15<r<18$) & 438 (34\%)
\\\\
gf1 & (\texttt{mgiant} or \texttt{bhb} or \texttt{rrl}) and ($13.5<r<17.5$) and ($\pi<0.3$)&
(\texttt{not} rank1) and ($\pi <0.3$) and ($15<r<18$) & 464 (37\%) \\\\
\enddata
\end{deluxetable*}

\begin{figure}[h!]
    \centering
    \includegraphics[width=0.45\textwidth]{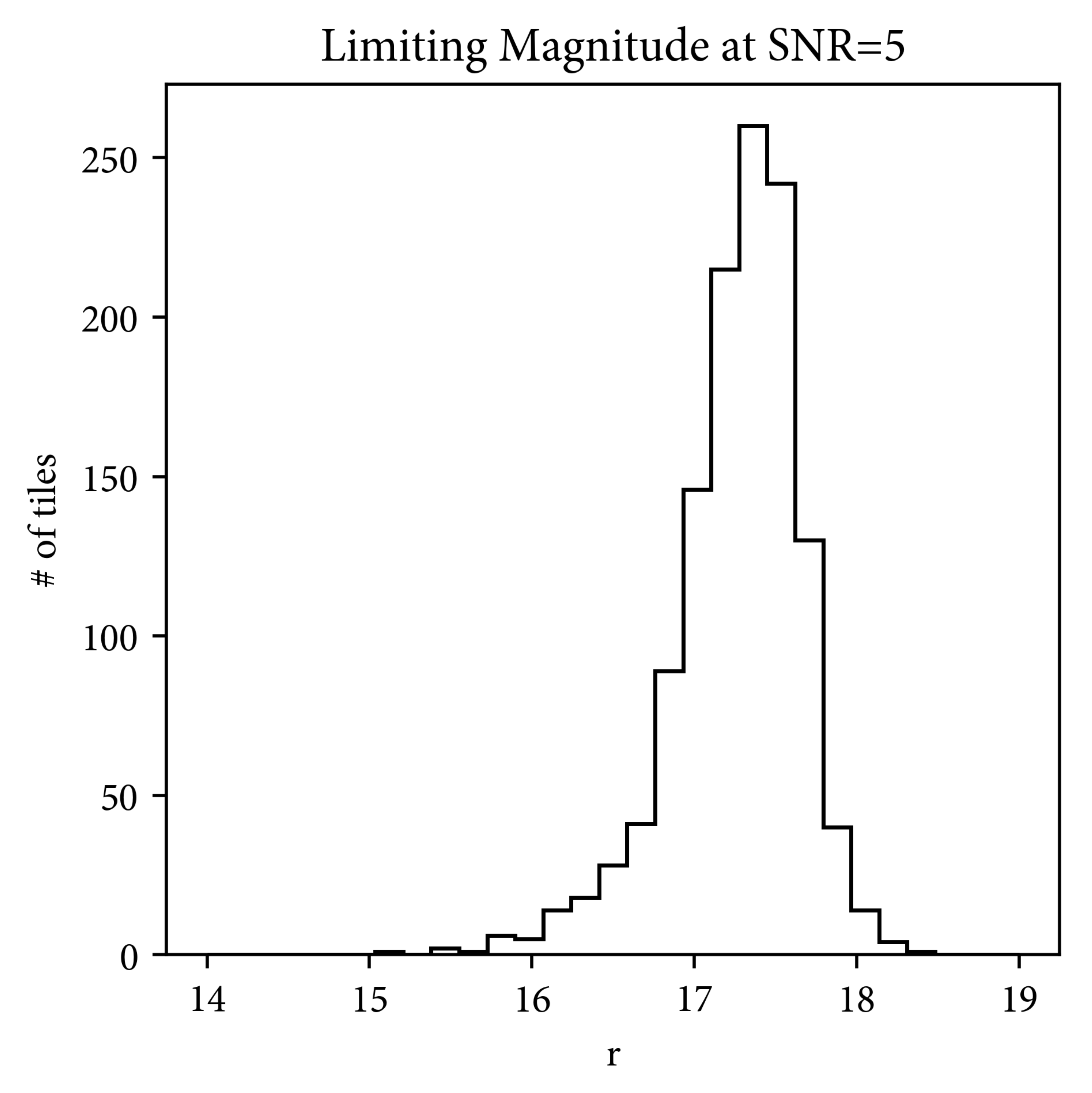}
    \caption{Limiting magnitude of the H3 Survey at a threshold of SNR=5. The main sample selection is $15<r<18$. Good observing conditions yield fainter limiting magnitudes, and bad weather can lead to brighter limiting magnitudes.}
    \label{fig:limitmag}
\end{figure}
\end{CJK*}
\end{document}